%% file: main.tex
\documentclass[11pt]{article}

\usepackage{graphicx} 
\usepackage{amsmath}
\usepackage{amsfonts}
\usepackage{amsthm}
\usepackage{bbm}
\usepackage{multirow}
\usepackage{array}
\usepackage[colorlinks=true, linkcolor=green, citecolor=blue, urlcolor=blue]{hyperref}
\usepackage{yfonts}
\usepackage{booktabs}
\usepackage{cleveref}
\usepackage[margin=1in]{geometry}
\usepackage{setspace}
\usepackage{comment}
\usepackage{todonotes}
\usepackage{amssymb}
\usepackage{color}
\usepackage{subcaption}
\usepackage{float}
\usepackage[round]{natbib}
\definecolor{vargreen}{RGB}{0,150,0}

\input{macros}

\theoremstyle{plain}
\newtheorem{theorem}{Theorem}[section]

\newtheorem{corollary}[theorem]{Corollary}
\theoremstyle{definition}

\newtheorem{remark}[theorem]{Remark}

\theoremstyle{remark}
\numberwithin{equation}{section}
\newtheorem*{empty*}{}

\title{Optimal Dynamic Fees in Automated Market Makers}

\author{Leonardo Baggiani\thanks{Corresponding author: leonardo.baggiani@warwick.ac.uk}\,\,\thanks{Department of Statistics, University of Warwick}
\and
Martin Herdegen\thanks{Department of Mathematics, University of Stuttgart}
	\and 
Leandro S\'anchez-Betancourt\thanks{Mathematical Institute and Oxford-Man Institute of Quantitative Finance, University of Oxford}
	}

\newcommand{\R}{\mathbb{R}}

\newcommand{\ra}[1]{\renewcommand{\arraystretch}{#1}}

\newcommand{\ind}{\mathbbm 1}

\newcommand{\dd}{\,\mathrm{d}}

\renewcommand{\epsilon}{\varepsilon}

\begin{document}

\maketitle

\begin{abstract}
    Automated Market Makers (AMMs) are emerging as a popular decentralised trading platform. A key recent development of the AMM technology is to allow for dynamic fees. In this work, we characterise the optimal dynamic fees in a constant function market maker. We find approximate closed-form solutions and study the optimal fee structure. We find that there are two distinct fee regimes: one in which the AMM imposes higher fees to deter arbitrageurs, and another where fees are lowered to increase volatility and attract noise traders. As price formation takes place in the AMM, these regimes disappear and the optimal fee structure becomes constant. Our results also show that dynamic fees that are linear in inventory and that track changes in the external price are a good approximation of the optimal fee structure and thus constitute suitable candidates when designing fees for AMMs. Within a stochastic pool depth framework, we show that the same qualitative fee regimes persist, with deeper pools charging lower fees.
    \newline

\vspace{0.7em}
\vspace{0.5em}

\vspace{0.5em}
\end{abstract}

\section{Introduction}

In recent years, automated market makers (AMMs) have gained popularity as a key innovation in financial technology, fundamentally changing the way liquidity is provided and exchanged within decentralized finance (DeFi) ecosystems. An AMM is a decentralized venue, where liquidity provision and liquidity taking follow rules that are encoded in smart contracts. As of 2025, the most popular AMM is Uniswap, which began gaining widespread adoption with its v2 version and has now reached v4; see \citet{Adams2020UniswapVC, Adams2021UniswapVC, Adams2024UniswapVC} for the white papers on versions v2 to v4, and \citet{Bachu2025} for a detailed overview of the v4 protocol.
\newline \newline
The early works of \citet{capponi2021adoptionblockchainbaseddecentralizedexchanges}, \citet{bartoletti2021theoryamm}, \citet{angeris2020priceoracles}, \citet{angeris2021uniswapanalysis}, \citet{Cartea_2022} and \citet{Angeris2023} first explain how these exchanges operate and introduce the problem of liquidity provision in AMMs. As shown by \citet{Cartea202369, Cartea2024931}, \citet{milionis2022automated}, and \citet{fukasawa2025liquidity}, liquidity provision in the absence of fees leads to losses for liquidity providers. There are two well-understood types of losses: \emph{loss-versus-rebalancing} (LVR) and \emph{impermanent loss} (IL). \cite{milionis2024automated} describe  LVR as
\begin{quote}
    \textit{the loss incurred by LPs relative to a trading strategy (the
``rebalancing strategy'') that holds the same risky positions as the pool, but that trades at market
prices rather than AMM prices.}
\end{quote} 
On the other hand, IL is the difference in value between holding the assets inside the pool and holding the assets outside of the pool.
The name \emph{impermanent} is due to the fact that if the market condition reverse,  
the losses disappear.
\newline \newline
To address the IL issue, the literature has generally taken two main approaches. The first is to redesign automated market makers (AMMs) to better account for spot price volatility. For instance, \citet{Bergault2024} propose an oracle-based AMM, where price discovery is not determined solely by liquidity takers. Instead, their pricing function integrates external information about the current market price of the risky asset.

The second approach focuses on finding trading strategies that allow liquidity providers to adapt to changing market conditions. \citet{Capponi2025} tackle the issue of IL by determining an optimal exit strategy for liquidity providers. \citet{Fukasawa03062023} address IL in constant function market makers by employing a weighted variance swap for hedging, and in a follow up study, \citet{fukasawa2024model} demonstrate a super-hedging strategy for IL that holds regardless of the centralised reference price dynamics. Most recently, \citet{sepp2024unified} show that IL can be fully hedged using European put and call options.
\newline\newline
To find solutions for the loss-versus-rebalancing problem, much of the research has centered on improving AMM designs to make liquidity provision less vulnerable to arbitrageurs. Several notable works in this direction include \citet{adams2025amammauctionmanagedautomatedmarket}, who introduce an auction-based AMM in which a pool “manager” captures arbitrage profits and rewards liquidity providers accordingly; \citet{goyal2023findingcurve}, who convert LPs’ beliefs about future asset values into optimal trading functions; and \citet{Cartea2024}, who generalize constant function market makers into decentralized liquidity pools (DLPs), enabling LPs to mitigate loss-versus-rebalancing by selecting suitable quote and impact functions and by monitoring external information. Another notable contribution is the work by \citet{nadkarni2024adaptive}, in which the authors investigate the problem of finding an optimal dynamic trading curve with the focus is specifically on developing a strategy that effectively hedges against loss-versus-rebalancing.
\newline \newline
As shown by \citet{milionis2024automated}, loss-versus-rebalancing can be partially or fully hedged through the collection of fees. Thus, the problem of determining optimal fees for AMMs has been an important issue since the early stages of research in this area. \citet{evans2021optimalfees} focused on how to find optimal fees for geometric mean market makers (G3Ms) while \citet{fritsch2021noteoptimalfees} studied how to determine optimal fees in a setting of multiple liquidity pools competing with each other for trade volume. \citet{he2024optimaldesignautomatedmarket} determined the optimal fee structure for a liquidity provider who allocates their capital between an automated market maker (AMM) and centralized trading venues. However, these works only address the problem of finding constant static fees. \citet{sabate2023variablefees} address the loss-versus-rebalancing problem by setting dynamic fees although no optimality is established.  \citet{Hasbrouck2022} found how higher fees on decentralised exchanges (DEX) can increase trading volume because higher fees attract more liquidity providers and boost the inventory in the DEX. Recently, \cite{aqsha2025equilibriumrewardliquidityproviders} solve a stochastic leader-follower game between the AMM and the liquidity providers in the venue. They find that under the equilibrium contract, the liquidity providers have incentives to add liquidity to the pool only when higher liquidity on average attracts more noise trading. In their work, the fee that the venue charges is proportional to the trade size.
\newline \newline
In this paper, we address the problem of determining optimal dynamic fees for AMMs.
As argued in \citet{Cao2023}, fees should adjust dynamically in order to reflect market conditions.
With the recent introduction of Uniswap v4, the implementation of dynamic fees is now possible.
Setting optimal fees is crucial because fees are the primary incentive for liquidity providers to engage with the AMM, as they help offset losses from impermanent loss.
As a result, a platform that optimizes fee collection can better compensate liquidity providers and, in turn, strengthen its overall market position.

Inspired by the market-making framework of \citet{zbMATH05278232},\footnote{For a comprehensive exposition of the market-making problem and its generalizations, see the monographs by \citet{zbMATH06459781} and \citet{zbMATH06550567}.}
we formulate a stochastic control problem in which the venue aims to maximize the total fees collected in a liquidity pool consisting of two assets: a risky asset $Y$ and a riskless asset $X$.

The control processes are $\feeLTsells$ and $\feeLTbuys$, representing the fees charged when liquidity takers sell and buy asset $Y$, respectively.\footnote{We choose $\feeLTsells$ for \emph{plus} and $\feeLTbuys$ for \emph{minus} because the quantity of asset $Y$ in the pool increases after a trade that pays $\feeLTsells$ and decreases after a trade that pays $\feeLTbuys$.}
These fees affect the exchange rate faced by liquidity takers: buying $Y$ requires paying more of asset $X$, while selling $Y$ yields less of asset $X$.
We assume that the order flow is generated by two types of market participants: noise traders and arbitrageurs.
Noise traders submit buy and sell orders for reasons that are independent of the fees set by the AMM.
Their order flow is therefore modeled through jump processes with exogenous intensities.
Arbitrageurs, instead, react to discrepancies between the AMM exchange rates and the price available on a centralized venue.

When selling asset $Y$ to the pool is sufficiently attractive relative to the centralized-exchange ask price, arbitrageurs submit sell orders to the AMM.
Conversely, when buying asset $Y$ from the pool is sufficiently attractive relative to the centralized-exchange bid price, they submit buy orders to the AMM.
In this way, arbitrageurs generate an order flow that depends on the controlled AMM exchange rates, as well as on the corresponding trade sizes. Rather than keeping noise traders and arbitrageurs as two separate populations throughout the analysis, we aggregate their effects into a representative liquidity taker. We therefore model sell and buy orders through two controlled point processes,
$\{\ppsell_t\}_{t \in [0,T]}$ and $\{\ppbuy_t\}_{t \in [0,T]}$,
whose intensities depend on the fees chosen by the AMM, the resulting exchange rates, and the state of the pool. The precise functional form of these intensities is introduced in the model section.

The control problem we wish to solve is, in its full generality, intractable. Therefore, we study two cases for which we obtain closed-form solutions. First, we take a vanishing viscosity approach and solve the problem treating $S_{t}$ as a parameter. Under this assumption, we derive in Theorem \ref{th: value function and optimal fees} the optimal fee policies and the corresponding value function.
Second, we approximate the exponential terms in the Hamilton–Jacobi–Bellman (HJB) equation using a linear-quadratic expansion. This leads to an alternative tractable formulation from which we derive the optimal fees and value function, as stated in Theorem \ref{thm: solution to the PDE system}.
In both approximations, we obtain the same insight: the venue finds it optimal to adopts two distinct fee regimes; one in which the AMM imposes higher fees to deter arbitrageurs, and another where fees are lowered to increase volatility and attract noise traders.

We run simulations to compare the performance of the optimal fee policies against alternative strategies. We find that the optimal fees outperform constant fee strategies by a significant margin. Furthermore, a linear approximation of the optimal policies proves to be sufficiently accurate, yielding nearly identical revenue outcomes.
Our results suggest that the optimal fees balance two competing objectives: setting fees high enough to maximize per-trade revenue while simultaneously increasing the quadratic variation of the marginal price, thereby attracting increased noise trading. A similar pattern emerges under the second approximation, where the revenue generated by the optimal fees also exceeds that of the alternative strategies. We then extend the model to allow for stochastic pool depth. This extension is motivated by the fact that liquidity in real AMMs is not fixed: liquidity providers may add or withdraw capital in response to market conditions, expected fee revenues, volatility, and outside opportunities. We model the pool depth on a finite grid and derive the corresponding dynamic programming equation and optimal fee rules. The stochastic-depth extension shows that the two-regime fee structure is robust to time-varying liquidity. It also highlights an additional effect: deeper pools optimally charge lower percentage fees, since larger trade sizes allow the venue to collect comparable fee revenues while remaining more competitive.

Finally, we use the optimal fee schedules to study the equilibrium liquidity level that is induced by the venue’s fee policy. Since the optimal controls depend on the pool depth, each initial liquidity level generates a different feedback fee schedule. We therefore define an equilibrium liquidity level by considering the response of liquidity providers who evaluate the fees received from the venue against the impermanent loss generated by their position. This provides a simple reduced-form link between the venue’s optimal fee policy and the amount of liquidity that the pool is able to attract.

The remainder of the paper is organised as follows. Section \ref{section: the model} introduces the model that we use, sets up the control problem and the performance criterion that the venue wants to maximize. In Section \ref{section: Optimal Strategy with Exponential Intensities} we compute the optimal policies. Sections \ref{section: First Approximation} and \ref{section: second Approximation} are dedicated to finding the value functions for the first and second approximation, respectively. In Sections \ref{Optimal fee structure: constant external price} and \ref{section: Optimal fee structure: dynamic external price with second order approximation} we carry out simulations to illustrate our theoretical results. Finally Section \ref{sec: Stochastic depth} extends the model to stochastic pool depth and uses the optimal fee schedules to characterize the equilibrium liquidity level induced by the venue’s fee policy.

\section{The model}\label{section: the model}
We study an automated market maker (AMM) with a finite time horizon $T > 0$. More precisely, we consider a constant function market (CFM) pool inside the reference AMM
with a riskless asset $X$ and a risky asset $Y$. Let $\depth > 0$ denote the depth of the pool and let $f: \mathbb{R}_{+} \times \mathbb{R}_{+} \to \mathbb{R}_{+}$ be the trading function, which is strictly increasing and twice differentiable
in both arguments. 
The trading condition is $f(x, y) = \depth$, where $x$ and $y$ denote the amounts in assets $X$ and $Y$, respectively. We can rewrite this condition using the level function $\varphi: \mathbb{R}_{+} \to \mathbb{R}_{+}$ satisfying $f(\varphi(y), y) = \depth$. The latter is then automatically differentiable and decreasing, and we assume in addition that it is convex in order to exclude roundtrip arbitrage, see \citet{Cartea2024931}.
These assumptions are satisfied by the most popular CFMs such as constant product markets (CPMs) where $f(x,y) = x y$ and $\varphi(y) = \depth/y$. 

\medskip{}
In our model, the quantity of asset $Y$ in the pool takes values in a finite grid given by
\begin{equation*}\label{eq: grid for quantity of y}
         \{ y^{-N} := \underline{y}, \dots, y^{0}, \dots, y^{N} := \overline{y} \},
\end{equation*}
where $\underline{y}$ and $\overline{y}$ can be interpreted as reserve constraints satisfying $0 <\underline{y} < y^{0} < \overline{y}$.

Consequently, the quantity of asset $X$ in the pool takes values in the grid 
\begin{equation*}\label{eq: grid for quantity of x}
        \{ x^{-N} :=\varphi(y^{-N}), \dots, x^{0} := \varphi(y^0), \dots, x^{N} := \varphi(y^{N}) \}.
\end{equation*}
Note that the monotonicity of the grids for the two assets are opposite to each other: the grid for asset $Y$ is increasing, whereas the grid for asset $X$ is decreasing. 

\medskip{}
There are three basic exchange rates for the price of asset $Y$ in units of asset $X$ in terms of the quantity of asset $Y$ in the pool that we use throughout the paper. To fix notation, let $y^{i}$ denote the quantity of asset $Y$ in the pool just before the trade of a liquidity taker.
\begin{enumerate}

    \item The \emph{marginal exchange rate} describes the price of an infinitesimal trade and is given by \begin{equation*}\label{eq: marginal exchange}
        Z ( y^{i} ) : = - \varphi' (y^{i}).
    \end{equation*}

    \item The \emph{exchange rate for buying} (taking out of the pool) $\deltabuy(y^{i}):= y^i-y^{i-1}$ units of asset $Y$ is given by
    \begin{equation*}\label{eq: exchange rate for buying}
        \exratebuy(y^{i}) :  = \frac{\varphi(y^{i-1}) - \varphi(y^{i})}{\deltabuy(y^{i})}.
    \end{equation*}
It takes values in  the grid
    \begin{equation*}\label{eq: grid for buying exchange}
      \{ \exratebuy^{-N + 1} := \exratebuy(y^{-N+1}), \dots, \exratebuy^{0} :=  \exratebuy(y^{0}), \dots, \exratebuy^{N} :=  \exratebuy(y^{N})\}.
    \end{equation*}
    \item The \emph{exchange rate for selling} (depositing in the pool) $\deltasell(y^{i}):= y^{i+1}-y^{i}$ units of asset $Y$ is given by
    \begin{equation*}\label{eq: exchange rate for selling}
        \exratesell(y^{i}) : = \frac{\varphi(y^{i}) - \varphi(y^{i+1})}{\deltasell(y^{i}) }.
    \end{equation*}
    It takes values in  the grid
    \begin{equation*}\label{eq: grid for selling exchange}
      \{ \exratesell^{-N} := \exratesell(y^{-N}), \dots, \exratesell^{0} := \exratesell(y^{0}), \dots, \exratesell^{N - 1} :=  \exratesell(y^{N-1}) \}.
    \end{equation*}
\end{enumerate}
Note that exchange rates for buying and selling satisfy the identity 
\begin{equation*}
\exratebuy^{i} = \exratesell^{i-1}, \quad i \in \{ -N + 1, \dots, 0, \dots N\}.
\end{equation*}  

We now add proportional transaction cost to the trading mechanism, and model the fees for buying and selling with controls $\feeLTbuys: \{ y^{-N}, \ldots, y^{N} \} \to \R$ and $\feeLTsells: \{ y^{-N}, \ldots, y^{N} \} \to \R$, respectively. This specification is consistent with recent AMM architectures such as Uniswap v4, where hooks allow pool creators to implement customised fee logic. In particular fees can be made state-dependent and may distinguish between trades that buy asset $Y$ from the pool and trades that sell asset $Y$ to the pool. Although one would expect that $\feeLTbuys,\feeLTsells \in [0,1]$, we do not impose this condition since it turns out to be optimal to sometimes charge negative fees, meaning that the venues pays liquidity takers to trade in a given direction.

The fees are collected in units of the riskless asset $X$ and are deposited outside of the pool. The corresponding (fee-depending) exchange rates can then be described in the following way, where $y^{i}$ denotes the quantity of asset $Y$ in the pool just before the trade.
 \begin{enumerate}
    \item The \emph{exchange rate with fee structure $\feeLTbuys$ for buying} (taking out of the pool) $\deltabuy(y^{i}):= y^i-y^{i-1}$ units of asset $Y$ is given by
    \begin{equation*}\label{eq: exchange rate for buying with fees}
        \exratebuy^\feeLTbuys(y^{i}) : = (1 + \feeLTbuys(y^{i})) \exratebuy(y^{i}).
    \end{equation*}
    \item The \emph{exchange rate with fee structure $\feeLTsells$ for selling} (depositing in the pool) $\deltasell(y^{i}):= y^{i+1}-y^{i}$ units of asset $y$ is given by
   \begin{equation*}\label{eq: exchange rate for selling with fees}
        \exratesell^\feeLTsells(y^{i}) : = (1 - \feeLTsells(y^i)) \exratesell(y^{i}) .
\end{equation*}
\end{enumerate}

In this paper the AMM mechanism works independently of the fee structure because the fees charged by the venue go to the venue's cash account to be redistributed at a later stage. In the pool, trading is carried out via the rates $\exratesell$ and $\exratebuy$ so that the depth $\depth$ does not change. However, from the point of view of the liquidity taker, the exchange rates include fees.

\medskip{}
Our model aims to study how the AMM sets fees dynamically in order to maximize revenue coming from the order flow. We fix a probability space $(\Omega, \mathcal{F}, \mathbb{F} = \{ \mathcal{F}_{t} \}_{t \in [0,T]}, \mathbb{P}^{\feeLTsells,\feeLTbuys})$ supporting all the processes we use below.\footnote{A rigorous definition of the probability space would involve a weak formulation of the control problem because the controls affect the intensities of the jump processes that inform the filtration. We leave this rigorous treatment out of the paper for readability. See Section 3.1 in \citet{Barucci2025} for a formal treatment of this issue.}
We assume that there are two types of market participants: noise traders and arbitrageurs. Noise traders submit orders mostly for exogenous reasons but their trading activity 
does depend on the fees set by the AMM. However, it is independent of the centralised reference price.
We model their sell and buy orders by two controlled point processes
\[
    \{N^{n+}_t\}_{t\in[0,T]}
    \qquad\text{and}\qquad
    \{N^{n-}_t\}_{t\in[0,T]},
\]
with controlled intensities
\[
\lambda_{t}^{n+,\feeLTsells} : = c^{n+} \exp(- \alpha^+\feeLTsells_t), 
\]
and 
\[
  \lambda_{t}^{n-,\feeLTbuys} : =c^{n-}\exp( - \alpha^-\feeLTbuys_t),
\]
where $c^{a+}$ and $c^{a-}$ measure the size of the noise traders' activity and $\alpha^+$ and $\alpha^-$ are sensitivity parameters.

Arbitrageurs, by contrast, react to discrepancies between the exchange rates offered by the AMM and the prices available on a centralised exchange. At time $t$, arbitrageurs sell asset $Y$ to the pool whenever the controlled AMM exchange rate for selling to the pool, denoted by $\exratesell^{\feeLTsells_t}(y_{t-}^{\feeLTsells,\feeLTbuys})$, is larger than the centralised-exchange ask price $\oraclepricestochastic + \zeta$. Similarly, they buy asset $Y$ from the pool whenever the controlled AMM exchange rate for buying from the pool, denoted by $\exratebuy^{\feeLTbuys_t}(y_{t-}^{\feeLTsells,\feeLTbuys})$, is smaller than the centralised-exchange bid price $\oraclepricestochastic - \zeta$. We model the sell and buy orders of arbitrageurs by two controlled point processes
\[
    \{N^{a+}_t\}_{t\in[0,T]}
    \qquad\text{and}\qquad
    \{N^{a-}_t\}_{t\in[0,T]},
\]
with controlled intensities
\[
   \lambda_{t}^{a+}
   :=
   c^{a+}
   \left(
        \exratesell^{\feeLTsells_t}(y_{t-}^{\feeLTsells,\feeLTbuys}) - (\oraclepricestochastic +\zeta) 
   \right)^+
   \deltasell(y_{t-}^{\feeLTsells,\feeLTbuys}),
\]
and
\[
   \lambda_{t}^{a-}
   :=
   c^{a-}
   \left(
        (\oraclepricestochastic - \zeta) - \exratebuy^{\feeLTbuys_t}(y_{t-}^{\feeLTsells,\feeLTbuys})
   \right)^+
   \deltabuy(y_{t-}^{\feeLTsells,\feeLTbuys}).
\]
Here, $c^{a+}$ and $c^{a-}$ measure the size of arbitrageur activity, while $\Delta^+_t$ and $\Delta^-_t$ denote the quantities of asset $Y$ traded when selling to and buying from the pool, respectively.

We assume that the noise traders are independent from the arbitrageurs so that we can combine the two point processes for selling into a single (controlled) point process $\{N^{+}_t\}_{t \in [0, T]}$ and the two point processes for buying into a single (controlled) point process $\{N^{-}_t\}_{t \in [0, T]}$ with intensities given by
\begin{align*}
\lambda^+_t
&=
\lambda^{n+}+\lambda^{a+}_t  \\
&=
c^{n+} \exp(- \alpha^+\feeLTsells_t)
+
c^{a+}
   \left(
        \exratesell^{\feeLTsells_t}(y_{t-}^{\feeLTsells,\feeLTbuys}) - (\oraclepricestochastic +\zeta) 
   \right)^+
   \deltasell(y_{t-}^{\feeLTsells,\feeLTbuys}),
\end{align*}
and
\begin{align*}
\lambda^-_t
&=
\lambda^{n-}+\lambda^{a-}_t  \\
&=
c^{n-}\exp( - \alpha^-\feeLTbuys_t)
+
c^{a-}
   \left(
        (\oraclepricestochastic - \zeta) - \exratebuy^{\feeLTbuys_t}(y_{t-}^{\feeLTsells,\feeLTbuys})
   \right)^+
   \deltabuy(y_{t-}^{\feeLTsells,\feeLTbuys}).
\end{align*}

For analytical tractability, we replace the above sum by one exponential. More precisely, we do the following approximation
\begin{align*}
\lambda^+_t
&=
c^{n+} \exp(- \alpha^+\feeLTsells_t)
+
c^{a+}
\left(
    \exratesell^{\feeLTsells_t}
    (y_{t-}^{\feeLTsells,\feeLTbuys})
    -
    (\oraclepricestochastic+\zeta)
\right)^+
\deltasell(y_{t-}^{\feeLTsells,\feeLTbuys})
\\
&\approx c^{n+} (1 - \alpha^+\feeLTsells_t) + c^{a+}
\left(
    \exratesell^{\feeLTsells_t}
    (y_{t-}^{\feeLTsells,\feeLTbuys})
    -
    (\oraclepricestochastic+\zeta)
\right)\deltasell(y_{t-}^{\feeLTsells,\feeLTbuys}) 
\end{align*}
and using that $ \exratesell^\feeLTsells(y_{t-}^{\feeLTsells,\feeLTbuys})  = (1 - \feeLTsells_t) \exratesell(y_{t-}^{\feeLTsells,\feeLTbuys})$ we have
\begin{align*}
\lambda^+_t
&\approx
c^{n+} \Big(1 -\alpha^+ \Big(1 - \frac{S_t + \zeta}{Z_+(y_{t-}^{\feeLTsells,\feeLTbuys})}\Big)\Big)
\exp\left(
    \frac{c^{a+} + \frac{c^{n+}\alpha^+}{\deltasell(y_{t-}^{\feeLTsells,\feeLTbuys}) Z_+(y_{t-}^{\feeLTsells,\feeLTbuys})}}{c^{n+} \Big(1 -\alpha^+ \Big(1 - \frac{S_t + \zeta}{Z_+(y_{t-}^{\feeLTsells,\feeLTbuys})}\Big)\Big)}
    \left(
        \exratesell^{\feeLTsells_t}
        (y_{t-}^{\feeLTsells,\feeLTbuys})
        -
        (\oraclepricestochastic+\zeta)
    \right)
    \deltasell(y_{t-}^{\feeLTsells,\feeLTbuys})
\right)
\\
&\approx:
\overline{\lambda}^{+}
\exp\left(
    k^+
    \left(
        \exratesell^{\feeLTsells_t}
        (y_{t-}^{\feeLTsells,\feeLTbuys})
        -
        (\oraclepricestochastic+\zeta)
    \right)
    \deltasell(y_{t-}^{\feeLTsells,\feeLTbuys})
\right),
\end{align*}
where
\[
    \overline{\lambda}^{+}:=c^{n+} \Big(1 -\alpha^+ \Big(1 - \frac{s_0 + \zeta}{Z_+(y_0)}\Big)\Big),
    \qquad
   k^+:=\frac{c^{a+} + \frac{c^{n+}\alpha^+}{\deltasell(y_0) Z_+(y_0)}}{c^{n+} \Big(1 -\alpha^+ \Big(1 - \frac{s_0 + \zeta}{Z_+(y_0)}\Big)\Big)}.
\]
Similarly,
\begin{align*}
\lambda^-_t
&=
c^{n-}\exp( - \alpha^-\feeLTbuys_t)
+
c^{a-}
\left(
    (\oraclepricestochastic-\zeta)
    -
    \exratebuy^{\feeLTbuys_t}
    (y_{t-}^{\feeLTsells,\feeLTbuys})
\right)^+
\deltabuy(y_{t-}^{\feeLTsells,\feeLTbuys})
\\
&\approx
c^{n-}(1 - \alpha^-\feeLTbuys_t)
+
c^{a-}
\left(
    (\oraclepricestochastic-\zeta)
    -
    \exratebuy^{\feeLTbuys_t}
    (y_{t-}^{\feeLTsells,\feeLTbuys})
\right)
\deltabuy(y_{t-}^{\feeLTsells,\feeLTbuys})
\\
&\approx
c^{n-} \Big(1 -\alpha^- \Big(\frac{S_t -\zeta}{Z_-(y_{t-}^{\feeLTsells,\feeLTbuys})}-1\Big)\Big)
\exp\left(
   \frac{c^{a-} + \frac{c^{n-}\alpha^-}{\deltabuy(y_{t-}^{\feeLTsells,\feeLTbuys}) Z_-(y_{t-}^{\feeLTsells,\feeLTbuys})}}{c^{n-} \Big(1 -\alpha^- \Big(\frac{S_t -\zeta}{Z_-(y_{t-}^{\feeLTsells,\feeLTbuys})}-1\Big)\Big)}
    \left(
        (\oraclepricestochastic-\zeta)
        -
        \exratebuy^{\feeLTbuys_t}
        (y_{t-}^{\feeLTsells,\feeLTbuys})
    \right)
    \deltabuy(y_{t-}^{\feeLTsells,\feeLTbuys})
\right)
\\
&\approx:
\overline{\lambda}^{-}
\exp\left(
    k^-
    \left(
        (\oraclepricestochastic-\zeta)
        -
        \exratebuy^{\feeLTbuys_t}
        (y_{t-}^{\feeLTsells,\feeLTbuys})
    \right)
    \deltabuy(y_{t-}^{\feeLTsells,\feeLTbuys})
\right),
\end{align*}
where
\[
    \overline{\lambda}^{-}:=c^{n+} \Big(1 -\alpha^- \Big(\frac{s_0 -\zeta}{Z_-(y_0)}-1\Big)\Big),
    \qquad
    k^-:=\frac{c^{a-} + \frac{c^{n-}\alpha^-}{\deltabuy(y_0) Z_-(y_0)}}{c^{n-} \Big(1 -\alpha^- \Big(\frac{S_0 -\zeta}{Z_-(y_0)}-1\Big)\Big)}.
\]
This exponential specification preserves the economic mechanism of the original arbitrage-based intensities: sell orders to the pool become more likely when the AMM selling rate exceeds the centralised-exchange ask, while buy orders from the pool become more likely when the AMM buying rate lies below the centralised-exchange bid. At the same time, the exponential form yields a tractable control problem. Henceforth, we work with a representative liquidity taker on each side of the pool, combining noise traders and arbitrageurs. Their aggregate sell and buy orders are described by the controlled point processes
\[
    \{N^+_t\}_{t\in[0,T]}
    \qquad\text{and}\qquad
    \{N^-_t\}_{t\in[0,T]},
\]
with intensities given by
\begin{align*}\label{eq: intensities for ppp}
 \lambda_{t}^{-,\feeLTbuys} &: = \intbuy \exp{  \left( -\expdecay^{-}   (\exratebuy^{\feeLTbuys_{t}}(Y_{t-}^{\feeLTsells,\feeLTbuys}) - (\oraclepricestochastic -\zeta)) \deltabuy(Y_{t-}^{\feeLTsells,\feeLTbuys}) \right) } \ind_{ \{ Y_{t-}^{\feeLTsells,\feeLTbuys} > \underline{y} \}}, \\  
   \lambda_{t}^{+,\feeLTsells} &: =  \intsell \exp{  \left( \expdecay^{+}( \exratesell^{\feeLTsells_t}(Y_{t-}^{\feeLTsells,\feeLTbuys})  - (\oraclepricestochastic + \zeta)) \deltasell(Y_{t-}^{\feeLTsells,\feeLTbuys}) \right) } \ind_{ \{ Y_{t-}^{\feeLTsells,\feeLTbuys} < \overline{y} \}}.
\end{align*}
Here, $\intbuy$ and $\intsell$ are some baseline intensities, $k^+$ and $K^-$ are an exponential decay parameter (similar to that in \citet{zbMATH05278232}), $\{\oraclepricestochastic\}_{t \in [0, T]}$ denotes the midprice in a centralised exchange (outside of the pool) of asset $Y$ in terms of the riskless asset $X$, the so-called \emph{centralised reference price}, and $\zeta > 0$ is the corresponding half-spread in the centralised exchange. We assume that $\{\oraclepricestochastic\}_{t \in [0, T]}$ has dynamics given by \[ \oraclepricestochastic = S_0 + \sigma W_{t}, \] where $\{ W_{t} \}_{\{t \in [0,T]\}}$ is a Brownian motion.
In order to make the notation simpler, we assume $k^{+} = k^{-} : = k$, we incorporate $e^{-k \zeta}$ into $\intsell$ and $\intbuy$ and formally set $\zeta :=0$.
\medskip{}
The cumulative fees $\{\Cash_{t}^{\feeLTsells,\feeLTbuys}\}_{t \in [0, T]}$ collected by the AMM are in turn given by \begin{equation*} \label{eq : SDE for holdings}
\Cash_{t}^{\feeLTsells,\feeLTbuys} = \int_0^t \feeLTsells_{u}\exratesell^{\feeLTsells_{u}}(Y_{u^{-}}^{\feeLTsells,\feeLTbuys}) \deltasell(Y_{u^{-}}^{\feeLTsells,\feeLTbuys}) \dd N_{u}^{+} + \int_0^t \feeLTbuys_{u}\exratebuy^{\feeLTbuys_{u}}(Y_{u^{-}}^{\feeLTsells,\feeLTbuys}) \deltabuy(Y_{u^{-}}^{\feeLTsells,\feeLTbuys})) \dd N_{u}^{-}, \quad \quad t \in [0,T].
\end{equation*}
The performance criterion of the venue is given by
\begin{equation*} \label{eq: goal function}
    J(\feeLTbuys,\feeLTsells) : =  \mathbb{E} \left[ \Cash_{T}^{\feeLTsells,\feeLTbuys} -  \int_{0}^{T} \Pi(Y_{t}^{\feeLTsells,\feeLTbuys},S_{t}) \dd t \right],
\end{equation*}
where $\Pi$ is a penalty function. An example of the function $\Pi$ that we consider below is the quadratic distance between the centralised reference price $S_{t}$ and the instantaneous exchange rate $Z(y_{t})$, i.e., \[\Pi(Y_{t}^{\feeLTsells,\feeLTbuys},S_{t}) = \pencons ( Z(Y_{t}^{\feeLTsells,\feeLTbuys}) - S_{t})^{2}. \]
Note, however, that $\Pi$ is not needed to add concavity to the problem, because the term $\Cash_T$ itself is concave in the controls. Our main results and insights are obtained when $\Pi = 0$.

\section{Characterization of the value function}\label{section: Optimal Strategy with Exponential Intensities}

The AMM seeks to solve the control problem \[ v(t,\cash,y,s) : = \sup_{(\feeLTbuys,\feeLTsells) \in \mathcal{A}_{t}} v^{(\feeLTbuys,\feeLTsells)}(t,\cash,y,s),\]
where $\mathcal{A}_{t}$ denotes the set of all $\mathbb{F}$-predictable and bounded fee structure processes $(\feeLTbuys_{u},\feeLTsells_{u})_{\{t \leq u \leq T\}}$ and the conditional performance criterion is given by \[ v^{(\feeLTbuys,\feeLTsells)}(t,\cash,y,s) : = \mathbb{E}_{(t,\cash,y,s)} \left[ \Cash_{T}^{(t,\cash,y,s,\feeLTbuys,\feeLTsells)} - \int_{t}^{T}  \Pi(Y^{(t,\cash,y,s,\feeLTbuys,\feeLTsells)}_{u},S_{u}^{(t,\cash,y,s)}) \dd u \right]. \]
Here, $\{ \Cash_{u}^{(t,\cash,y,s,\feeLTbuys,\feeLTsells)} \}_{u \in [t,T]}$, $\{ Y_{u}^{(t,\cash,y,s,\feeLTbuys,\feeLTsells)} \}_{u \in [t,T]}$ and $\{ S_{u}^{(t,\cash,y,s)}\}_{u \in [t,T]} $ denote the (controlled) processes $\Cash$, $y$ and $S$ restarted at time $t$ with initial value $\cash$, $y$ and $s$, respectively.
From the dynamic programming principle, we determine that the Hamilton-Jacobi-Bellman (HJB) equation satisfied by the value function is
{\scriptsize
\begin{align} \label{eq: real HJB for v}
    & \frac{\partial}{\partial t} v (t,y,\cash,s) + \frac{\sigma^{2}}{2} \frac{\partial^{2}}{\partial s^{2}} v(t,y,\cash,s) - \Pi(y,s) \nonumber \\
    & +\left( \intsell e^{\expdecay(\exratesell( y)- \oracleprice) \deltasell(y)} \sup_{ \feeLTsells  \in \mathbb{R}} e^{-\expdecay \feeLTsells\exratesell( y) \deltasell(y)} \left[v(t,y + \deltasell(y) ,\cash + \feeLTsells \exratesell(y) \deltasell(y),s) - v(t,\cash,y,s)\right] \right) \ind_{ \{ y < \overline{y} \}} \\
    & +\left(  \intbuy e^{-\expdecay(\exratebuy(y)- \oracleprice) \deltabuy(y)} \sup_{ \feeLTbuys \in \mathbb{R}} e^{-\expdecay \feeLTbuys \exratebuy(y) \deltabuy(y)} \left[v(t,y - \deltabuy(y),\cash +  \feeLTbuys \exratebuy(y) \deltabuy(y),s) - v(t,\cash,y,s) \right] \right) \ind_{ \{ y > \underline{y} \}} = 0, \nonumber
\end{align}}with terminal condition $v(T,y,\cash,s) = \cash$. The terms in this HJB equation have an intuitive interpretation. The terms 
\[ v(t,y + \deltasell(y) ,\cash + \feeLTsells \exratesell(y) \deltasell(y)) - v(t,y,\cash,s), \quad \quad v(t,y - \deltabuy(y),\cash +  \feeLTbuys \exratebuy(y) \deltabuy(y)) - v(t,y,\cash,s), \] 
represent the difference in the value function before and after a trade. If there is a sell trade, the quantity of asset $y$ in the pools jumps from $y$ to $y + \deltasell(y)$, and the AMM collects fees that amount to $\feeLTsells \, \exratesell(y) \deltasell(y)$, i.e., the percentage of fees $\feeLTsells$ times the quantity traded $\deltasell(y)$ times the sell exchange rate at the moment of the trade $\exratesell(y)$. Symmetrically, when there is a buy trade, the grid for the risky asset moves from $y$ to $y - \deltabuy(y)$, and the AMM collects fees that amount to $\feeLTbuys \, \exratebuy(y) \deltabuy(y)$. Each of the differences is multiplied by the respective intensity of order arrival.

Next, we make the ansatz $v(t,y,\cash,s) = g(t,y,s) + \cash$, to obtain that $g(t,y,s)$ satisfies the HJB equation
{\scriptsize
\begin{align}\label{eq: HJB for g}
    & \frac{\partial}{\partial t}  g(t,y,s) + \frac{\sigma^{2}}{2} \frac{\partial^{2}}{\partial s^{2}} g(t,y,s) - \Pi(y,s) \nonumber \\
    &+ \left( \intsell e^{\expdecay (\exratesell(y) - \oracleprice ) \deltasell(y)} \sup_{ \feeLTsells \in \mathbb{R}} e^{-\expdecay \feeLTsells \exratesell(y) \deltasell(y)} \left[g(t,y + \deltasell(y),s) - g(t,y,s) + \feeLTsells \exratesell(y) \deltasell(y)\right] \right) \ind_{ \{ y < \overline{y} \}}  \\
    &+ \left(  \intbuy e^{-\expdecay(
    \exratebuy(y) -\oracleprice) \deltabuy(y)} \sup_{ \feeLTbuys \in \mathbb{R}} e^{-\expdecay \feeLTbuys \exratebuy(y) \deltabuy(y)} \left[g(t,y - \deltabuy(y),s) - g(t,y,s) +  \feeLTbuys \exratebuy(y) \deltabuy(y) \right] \right) \ind_{ \{ y > \underline{y} \}} = 0, \nonumber
\end{align}}
with terminal condition $g(T,y,s) = 0$. The corresponding maximizers are given by
\begin{equation}\label{eq: maximizers}
\begin{aligned}
    \feeLTsells^{*}(t,y) & := \frac{g(t,y,s) - g(t,y + \deltasell(y),s)}{ \exratesell(y) \deltasell(y)} + \frac{1}{\expdecay \exratesell(y) \deltasell(y)}, \\
    \feeLTbuys^{*}(t,y) & := \frac{g(t,y,s) - g(t,y - \deltabuy(y),s)}{ \exratebuy(y) \deltabuy(y)} + \frac{1}{\expdecay \exratebuy(y) \deltabuy(y)}.
\end{aligned}
\end{equation}
\newline
Plugging the maximizers in \eqref{eq: HJB for g}, we obtain
\begin{align}\label{eq: eq for g}
    & \frac{\partial}{\partial t}  g(t,y,s) + \frac{\sigma^{2}}{2} \frac{\partial^{2}}{\partial s^{2}} g(t,y,s) - \Pi(y,s)  \nonumber \\
    & \quad + \left( \frac{\intsell e^{- \expdecay \oracleprice \deltasell(y) - 1}}{\expdecay} e^{\expdecay \exratesell (y)  \deltasell(y)} e^{ \expdecay ( g(t,y + \deltasell(y),s) - g(t,y,s) )} \right) \ind_{ \{ y < \overline{y} \}}  \\
    &\quad + \left( \frac{\intbuy e^{ \expdecay \oracleprice \deltabuy(y) - 1}}{\expdecay} e^{-\expdecay \exratebuy (y)  \deltabuy(y)} e^{ \expdecay ( g(t,y - \deltabuy(y),s) - g(t,y,s)} \right) \ind_{ \{ y > \underline{y} \}} = 0. \nonumber
\end{align}

\section{Approximate solution with vanishing viscosity}\label{section: First Approximation}

To obtain a first tractable approximation, we exploit the short time horizon of the problem and consider the limiting regime in which the diffusion component of the centralised reference price is negligible. This simplification is used only to obtain analytical insight into the structure of the optimal fees. A similar approach has been adopted in option market-making models with volatility arbitrage; see \cite{LucicTse2025OptionMarketMaking}.
\newline
All numerical simulations are carried out with a stochastic centralised reference price process, and the robustness of the approximation is assessed in Section \ref{section: second Approximation}, where we consider an alternative approximation that allows us to treat the full HJB equation.
\newline \newline
With this assumption in place and applying Cole-Hopf transform $e^{kg(t,y,s)} = w(t,y,s)$, the HJB equation \eqref{eq: eq for g} simplifies to
\begin{align} \label{eq: eq for w}
    & \frac{\partial}{\partial t}  w(t,y,s) - \expdecay \Pi(y,s) w(t,y,s)\nonumber \\
    &  +\left( \intsell e^{- \expdecay \oracleprice \deltasell(y) - 1} e^{\expdecay \exratesell (y)  \deltasell(y)} w(t,y + \deltasell(y),s) \right) \ind_{ \{ y < \overline{y} \}}  \\
    & +\left( \intbuy e^{ \expdecay \oracleprice \deltabuy(y) - 1} e^{-\expdecay \exratebuy (y)  \deltabuy(y)} w(t,y - \deltabuy(y),s) \right) \ind_{ \{ y > \underline{y} \}} = 0. \nonumber
\end{align}
This can be interpreted as a system of ODEs, parameterised by $s$. This can be solved by matrix exponentials, parametrised by $s$. The proof of the following result 
 is straightforward and hence, omitted.

\begin{theorem} \label{th: value function and optimal fees}
   For $s \in \mathbb{R}$ Define the matrix $ \mathbf{A}(s) : =  ( \mathbf{A}_{i,j}(s))_{0 \leq i \leq j \leq 2N} $ by
    \[
    \mathbf{A}_{i,j}(s) : =
    \begin{cases}
        - \expdecay \Pi(y^{j - N},s)  & \text{ if } i = j, \\
        \intsell e^{- \expdecay \oracleprice \deltasell(y^{j - N}) - 1} e^{\expdecay \exratesell (y^{j - N})  \deltasell(y^{j - N})} & \text{ if } i = j - 1, \\
        \intbuy e^{ \expdecay \oracleprice \deltabuy(y^{j - N}) - 1}e^{-\expdecay \exratebuy (y^{j - N})  \deltabuy(y^{j - N})} & \text{ if } i = j + 1 \\
        0 & \text{ otherwise,}
    \end{cases}
    \]
    and denote with $\mathbf{1}$ the unit vector of $\mathbb{R}^{2N}$. Define the function $w : [0,T] \times \{ y^{-N}, \dots, y^{N} \} \times \mathbb{R} \to \mathbb{R}$ by \[ w(t,y^{i},s) : = (\exp( \mathbf{A(s)}(T - t)) \mathbf{1})_{i}, \]
    and the function $v : [0,T] \times \{ y^{-N}, \dots, y^{N} \}  \times \mathbb{R} \times \mathbb{R} \to \mathbb{R}$ as
    \[ v(t,y^{i}, \cash, s) = \cash + \frac{1}{k}\log(w(t,y^{i},s)).\]
    Then $v$ solves the Hamilton-Jacobi-Bellman equation
\begin{align*}
    & \frac{\partial}{\partial t} v(t,y^{i},\cash,s) - \Pi(y^{i},s) + \nonumber \\
    & \left( \intsell e^{- \expdecay \oracleprice \deltasell(y^{i})} \sup_{ \feeLTsells \in \mathbb{R}} e^{ \expdecay \exratesell^{\feeLTsells}(y^{i}) \deltasell(y^{i})} \left[v(t,y^{i+1}, \cash + \feeLTsells \exratesell(y^{i}) \deltasell(y^{i}),s) - v(t,y^{i},\cash,s)\right] \right) \ind_{ \{ y^{i} < y^{N} \}} \\
    & \left(  \intbuy e^{\expdecay \oracleprice \deltabuy(y^{i})} \sup_{ \feeLTbuys \in \mathbb{R}} e^{-\expdecay \exratebuy^{\feeLTbuys}(y^{i}) \deltabuy(y^{i})} \left[v(t,y^{i-1},\cash +  \feeLTbuys \exratebuy(y^{i}) \deltabuy(y^{i}),s) - v(t,y^{i},\cash,s) \right] \right) \ind_{ \{ y^{i} > y^{-N} \}} = 0, \nonumber
\end{align*}
with boundary condition $v(T,y,\cash, s) = \cash$.  Moreover, the corresponding optimizers are independent of $\cash$ and satisfy

\begin{align}
    \feeLTsells^{*}(t,y^i,s) & : =  \frac{1}{\expdecay \exratesell(y^{i}) \deltasell(y^{i})} \left( 1 +  \log \left( \frac{w(t,y^{i},s)}{w(t,y^{i + 1},s)} \right) \right), \\
    \feeLTbuys^{*}(t,y^i,s) & : = \frac{1}{\expdecay \exratebuy(y^{i}) \deltabuy(y^{i})} \left( 1 +  \log \left( \frac{w(t,y^{i},s)}{w(t,y^{i - 1},s)} \right) \right),
\end{align}
for $y^{i} \in \{ y^{-N} = \underline{y}, \dots, y^{0} = y_{0}, \dots, y^{N} = \overline{y} \}$.
\end{theorem}

To construct a tractable implementation of the optimal fee rule, we also consider a first-order approximation of the optimal fees around the midpoint \(y^0\). Indeed, computing the optimal policy on the full grid requires evaluating a matrix exponential and then storing the resulting fee values at every inventory level. This can be computationally expensive, especially in larger state spaces, and may also lead to high implementation costs on-chain. A linear approximation provides a simpler alternative: it only requires evaluating the value function at a small number of points around \(y^0\), while still capturing the local behaviour of the optimal fee rule.

More precisely, we define
\[
\feeLTsells^{\mathrm{lin}}(t,y,s)
=
\feeLTsells^*(t,y^0,s)
+
\frac{\feeLTsells^*(t,y^1,s)-\feeLTsells^*(t,y^{-1},s)}{y^1-y^{-1}}
(y-y^0),
\]
and
\[
\feeLTbuys^{\mathrm{lin}}(t,y)
=
\feeLTbuys^*(t,y^0)
+
\frac{\feeLTbuys^*(t,y^1)-\feeLTbuys^*(t,y^{-1})}{y^1-y^{-1}}
(y-y^0).
\]
Using the explicit expressions in Theorem~\ref{th: value function and optimal fees}, this yields
\[
\feeLTsells^{\mathrm{lin}}(t,y,s)
=
\frac{
1+\log\left(\frac{w(t,y^0,s)}{w(t,y^1,s)}\right)
}{
kZ_+(y^0)\Delta_+(y^0)
}
+
\beta_{\feeLTsells}(t,s)(y-y^0),
\]
where
\[
\beta_{\feeLTsells}(t,s)
=
\frac{1}{y^1-y^{-1}}
\left[
\frac{
1+\log\left(\frac{w(t,y^1,s)}{w(t,y^2,s)}\right)
}{
kZ_+(y^1)\Delta_+(y^1)
}
-
\frac{
1+\log\left(\frac{w(t,y^{-1},s)}{w(t,y^0,s)}\right)
}{
kZ_+(y^{-1})\Delta_+(y^{-1})
}
\right],
\]
and
\[
\feeLTbuys^{\mathrm{lin}}(t,y,s)
=
\frac{
1+\log\left(\frac{w(t,y^0,s)}{w(t,y^{-1},s)}\right)
}{
kZ_-(y^0)\Delta_-(y^0)
}
+
\beta_{\feeLTbuys}(t)(y-y^0),
\]
where
\[
\beta_{\feeLTbuys}(t,s)
=
\frac{1}{y^1-y^{-1}}
\left[
\frac{
1+\log\left(\frac{w(t,y^1,s)}{w(t,y^0,s)}\right)
}{
kZ_-(y^1)\Delta_-(y^1)
}
-
\frac{
1+\log\left(\frac{w(t,y^{-1},s)}{w(t,y^{-2},s)}\right)
}{
kZ_-(y^{-1})\Delta_-(y^{-1})
}
\right].
\]
This approximation preserves the local level and slope of the optimal fee functions around the midpoint \(y^0\), while reducing the implementation to a linear inventory-dependent rule.

\subsection{Optimal fee structure}\label{Optimal fee structure: constant external price}

\begin{figure}[H]
    \centering
    \begin{subfigure}[b]{0.49\textwidth}
        \centering
        \includegraphics[width=\textwidth]{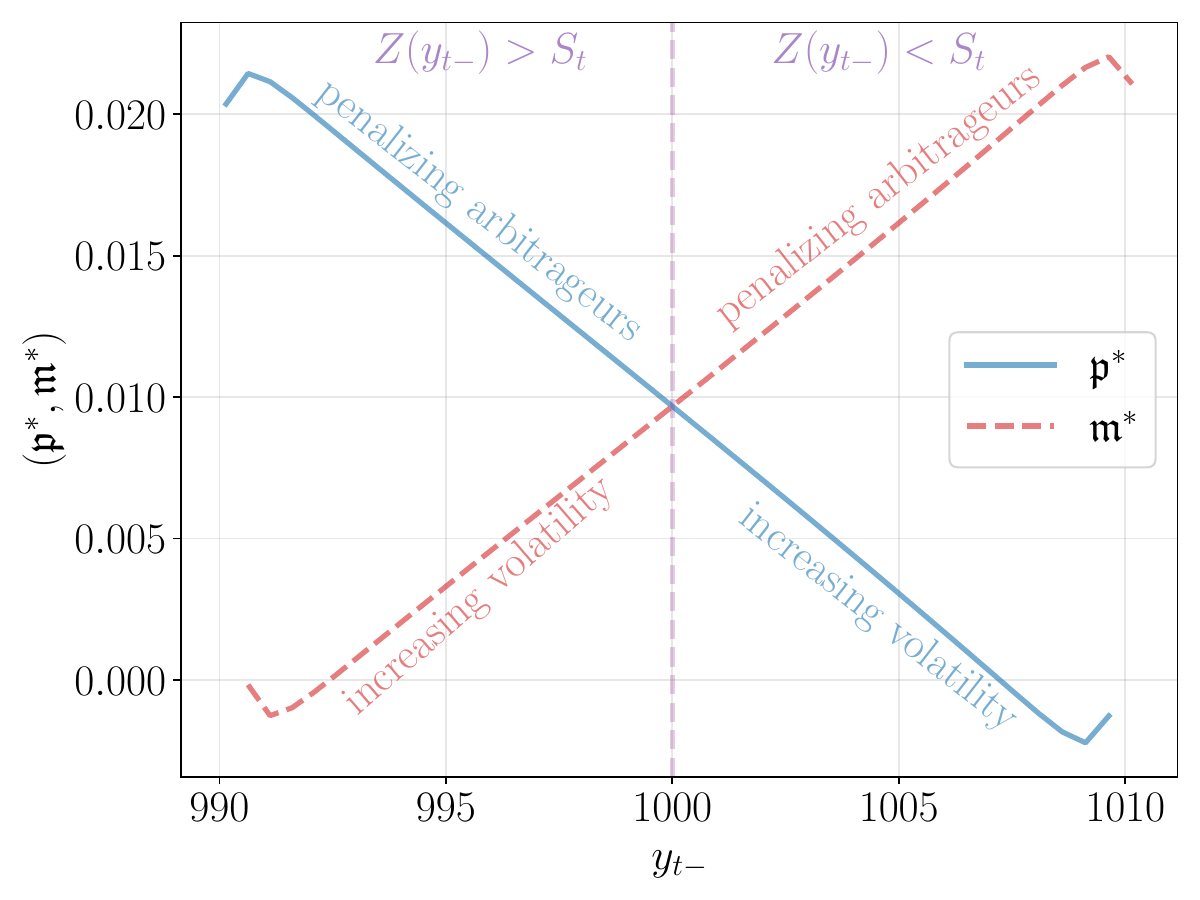}
        \caption{Optimal fees for selling $\feeLTsells^{*}(t,y_{t-},s)$ (solid line) and for buying $\feeLTbuys^{*}(t,y_{t-},s)$ (dashed line) at time $t=0.5$ and $s = 100$ as a function of the quantity of asset $Y$ in the pool.}
        \label{fig: optimal fees}
    \end{subfigure}
    \hfill
    \begin{subfigure}[b]{0.49\textwidth}
        \centering
        \includegraphics[width=\textwidth]{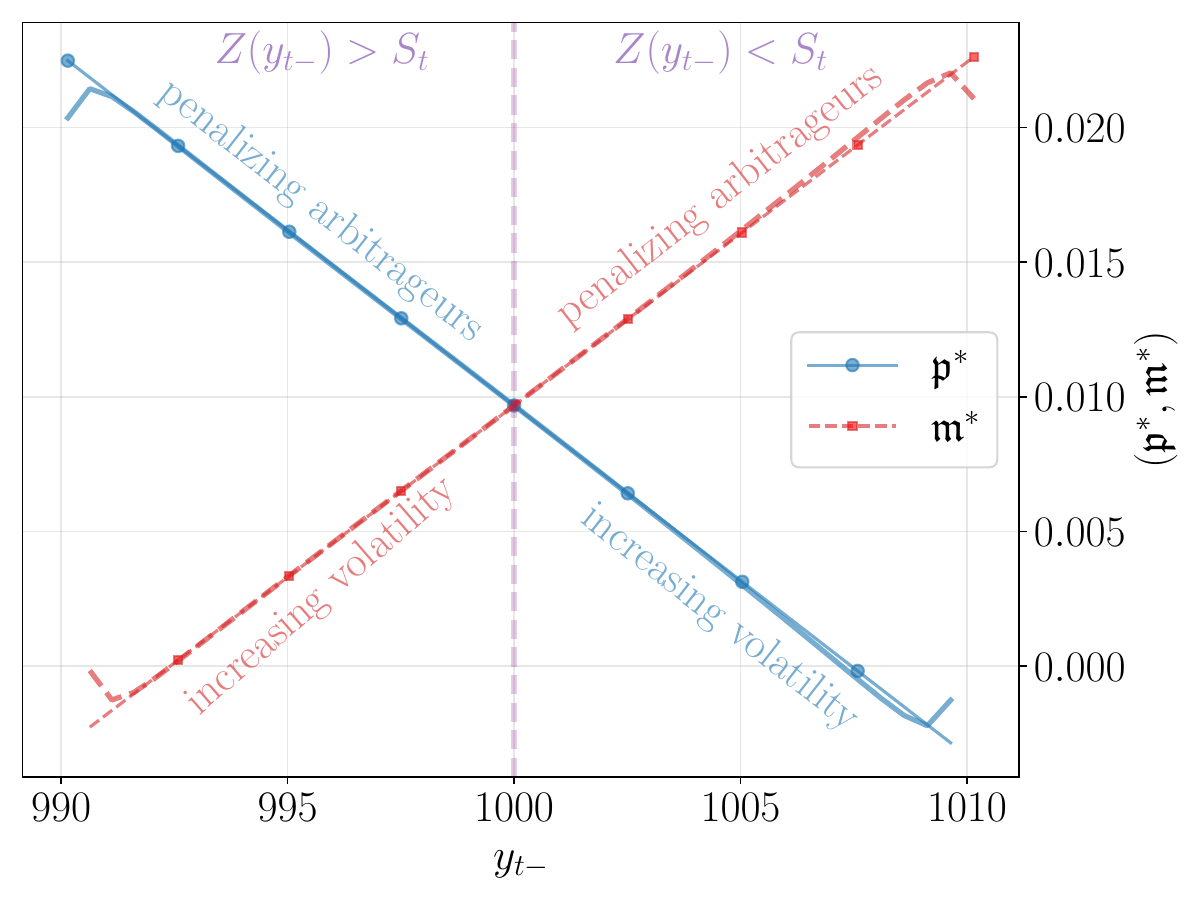}
        \caption{Linear fees for selling $\feeLTsells^{*}(t,y_{t-},s)$ (solid line) and for buying $\feeLTbuys^{*}(t,y_{t-},s)$ (dashed line) at time $t=0.5$ and $s = 100$ as a function of the quantity of asset $Y$ in the pool.}
        \label{fig: linear fees}
    \end{subfigure}
\end{figure}

We perform numerical simulations of the results found in Theorem \ref{th: value function and optimal fees}. We assume that the AMM is a CPM, where the trading function $f: \mathbb{R}_{+} \times \mathbb{R}_{+} \to \mathbb{R}_{+}$ is $f(x,y) := xy$ and the level function $\varphi: \mathbb{R}_{+} \to \mathbb{R}_{+}$ is $\varphi(y) := \depth / y$. We assume that the the grid for $y$ is such that after every trade the exchange rate $Z(y)$ moves by $0.1$. More precisely, the grid for the risky asset is in turn given by the formula \[ y^{i} : = \sqrt{\frac{\depth}{ \left( Z(y_0) - 0.1\,i  \right) } }, \quad \quad \text{ for } i \in \{ -20, \dots, 20 \}. \]
We take the time horizon $T=1$, the two baseline intensities $\intbuy = \intsell = 50$, the rate of exponential decay $\expdecay = 2$, the centralised reference price $S_0 =100$, and the penalty function $\Pi(y,s) = \pencons(Z(y) -s)^{2}$. Moreover, we assume that the initial conditions of the pool are $\depth = 10^{8}$ and $y_0 = 1000$. Figure \ref{fig: optimal fees} shows the optimal fees in the case where the penalization constant $\pencons = 0$.

From the plot in Figure \ref{fig: optimal fees} we see that the AMM charges fees in two different regimes. When the quantity of asset $y$ is below the value of $y_0$ it is more profitable to sell inside the AMM with respect to the centralised reference price, hence the AMM will charge high fees to penalize arbitrageurs. At the same time it will charge low fees to buy, even negative ones, in order to encourage noise traders to trade inside the venue and increase the volatility in the price. Symmetrically, the analogous happens when the quantity is above the value of $y_0$. Figure \ref{fig: linear fees} shows that if the quantity of asset $Y$ stays in a small enough neighbourhood of $y_0$, a linear approximation of the fees is a good enough approximation. Below we show that the linear approximation generates revenues close to the optimal strategy.

\begin{figure}[h]
    \centering
    \begin{subfigure}[b]{0.49\textwidth}
        \centering
        \includegraphics[width=\textwidth]{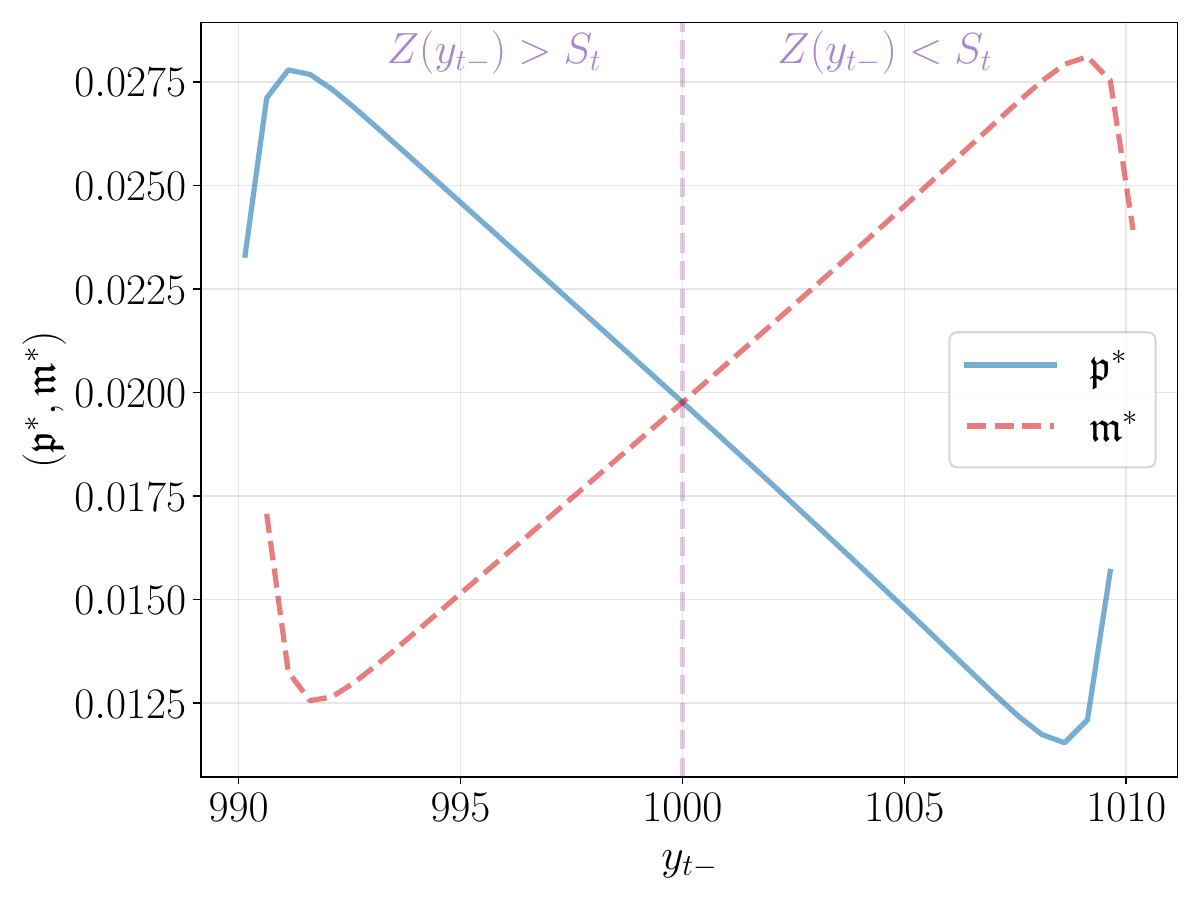}
        \caption{Optimal fees for selling $\feeLTsells^{*}(t,y_{t-},s)$ (solid line) and for buying $\feeLTbuys^{*}(t,y_{t-},s)$ (dashed line) multiplied by $k$ at time $t=0.5$ and $s = 100$ as a function of the quantity of asset $Y$ in the pool when $k = 1$.}
        \label{fig: optimal fees k=0.5}
    \end{subfigure}
    \hfill
    \begin{subfigure}[b]{0.49\textwidth}
        \centering
        \includegraphics[width=\textwidth]{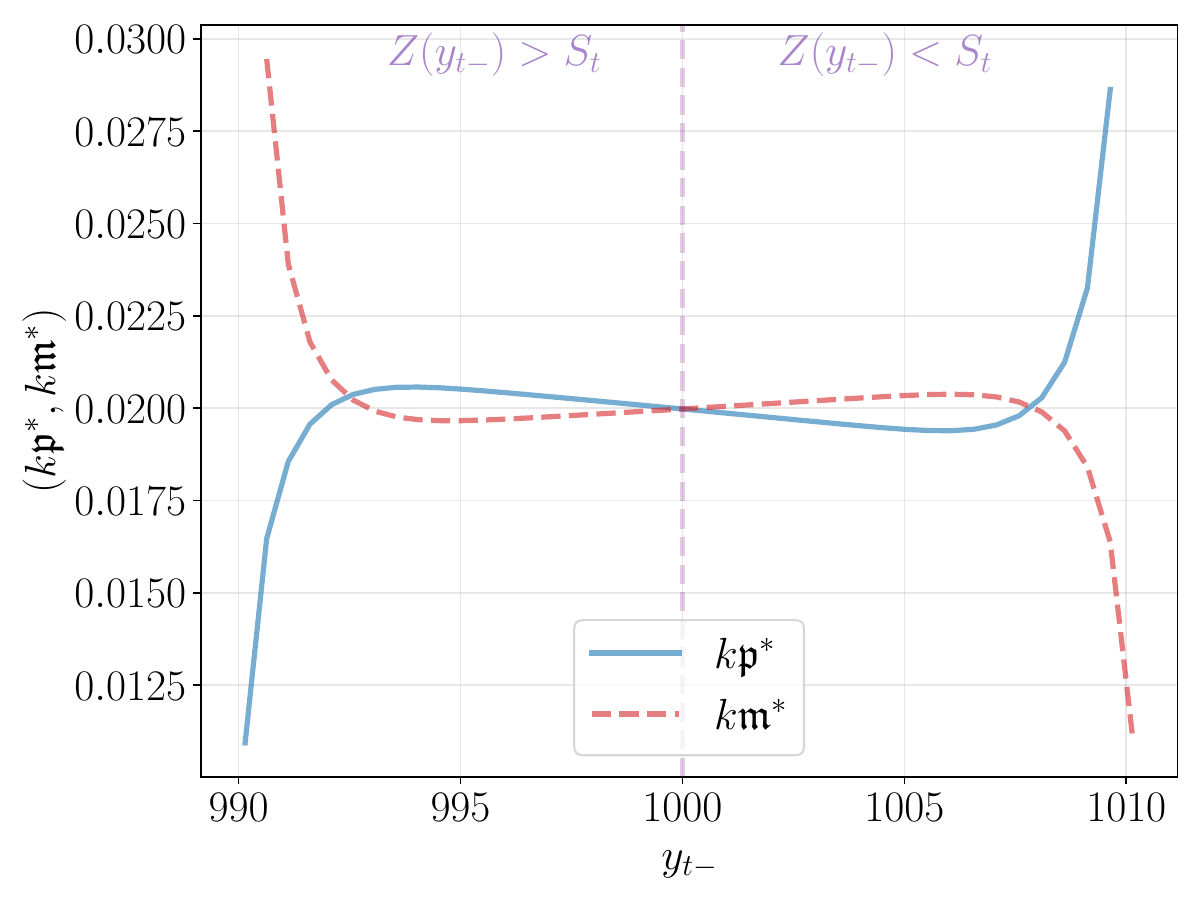}
        \caption{Optimal fees for selling $\feeLTsells^{*}(t,y_{t-},s)$ (solid line) and for buying $\feeLTbuys^{*}(t,y_{t-},s)$ (dashed line) multiplied by $k$ at time $t=0.5$ and $s = 100$ as a function of the quantity of asset $Y$ in the pool when $k = 0.25$.}
        \label{fig: optimal fees k=0.25}
    \end{subfigure}
\end{figure}

The behaviour of the fees is highly dependent on the value of the parameter $k$. Indeed, as the value of the parameter $k$ gets smaller, the probability of arrival becomes less sensitive to price changes. As a response to this, the AMM prioritizes pushing the price as close to the centralised reference price as possible to get away from the boundaries where it does not generate revenues. We can observe this phenomenon in Figure \ref{fig: optimal fees k=0.5} and Figure \ref{fig: optimal fees k=0.25} for $k=1$ and $k=0.25$, respectively.

The following corollary gives a rigorous mathematical characterization of this behaviour. The proof is straightforward and hence, omitted.

\begin{corollary}\label{cor: fee for k to 0}
    Define the matrix $ \mathbf{A} : =  ( \mathbf{A}_{i,j} )_{0 \leq i \leq j \leq 2N} $ by
    \[
    \mathbf{A}_{i,j} : =
    \begin{cases}
        \intsell e^{-1} & \text{ if } i = j - 1, \\
        \intbuy e^{-1} & \text{ if } i = j + 1, \\
        0 & \text{ otherwise,}
    \end{cases}
    \]
    and denote with $\mathbf{1}$ the unit vector of $\mathbb{R}^{2N}$. Define the function $w : [0,T] \times \{ y^{-N}, \dots, y^{N} \} \to \mathbb{R}$ by \[ w(t,y^{i}) : = (\exp( \mathbf{A}(T - t)) \mathbf{1})_{i},. \]
Let $\feeLTsells^{*}$ and $\feeLTbuys^{*}$ be the optimal fees from Theorem \ref{th: value function and optimal fees} and define the quantities
    \begin{align}
    \feeLTsells^{*}_0(t,y^i) & : =  \frac{1}{ \exratesell(y^{i}) \deltasell(y^{i})} \left( 1 +  \log \left( \frac{w(t,y^{i})}{w(t,y^{i + 1})} \right) \right) \\
    \feeLTbuys^{*}_0(t,y^i) & : = \frac{1}{ \exratebuy(y^{i}) \deltabuy(y^{i})} \left( 1 +  \log \left( \frac{w(t,y^{i})}{w(t,y^{i - 1})} \right) \right).
\end{align}
We have that 
\begin{align}
    \lim_{k \to 0} k\feeLTsells^{*}(t,y^i,s) = \feeLTsells^{*}_0(t,y^i) \quad \quad \text{ and } \quad \quad  \lim_{k \to 0} k\feeLTbuys^{*}(t,y^i,s) = \feeLTbuys^{*}_0(t,y^i),
\end{align}
for every $t \in [0,T]$, $y^{i} \in \{ y^{-N}, \dots, y^{N} \}$ and $s \in \mathbb{R}$.
\end{corollary}

The next plots illustrates the corollary. More precisely, we compare $k\feeLTsells^{*}$ and $k\feeLTbuys^{*}$ for $k=0.1$ (left figure) with the limit optimal fees $\feeLTsells^{*}_0$ and $\feeLTbuys^{*}_0$ (right figure) in \eqref{cor: fee for k to 0}.
\begin{figure}[H]
    \centering
    \begin{subfigure}[t]{0.49\textwidth}
        \centering
        \includegraphics[width=\textwidth]{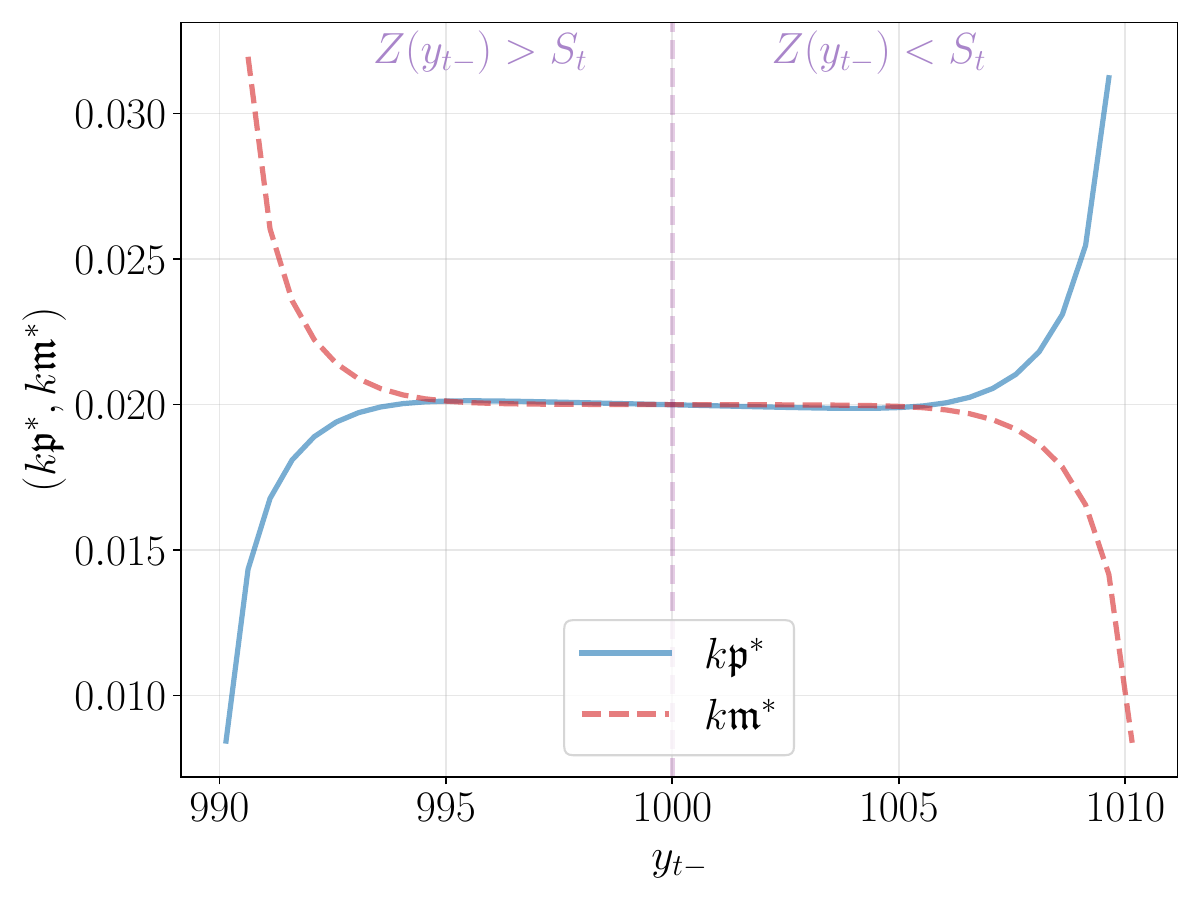}
        \caption{Optimal fees for selling $\feeLTsells^{*}(t,y_{t-},s)$ (solid line) and for buying $\feeLTbuys^{*}(t,y_{t-},s)$ (dashed line) multiplied by $k$ at time $t=0.5$ and $s = 100$ as a function of the quantity of asset $Y$ in the pool when $k = 0.5$.}
        \label{fig: optimal fees k=0.1}
    \end{subfigure}
    \hfill
    \begin{subfigure}[t]{0.49\textwidth}
        \centering
        \includegraphics[width=\textwidth]{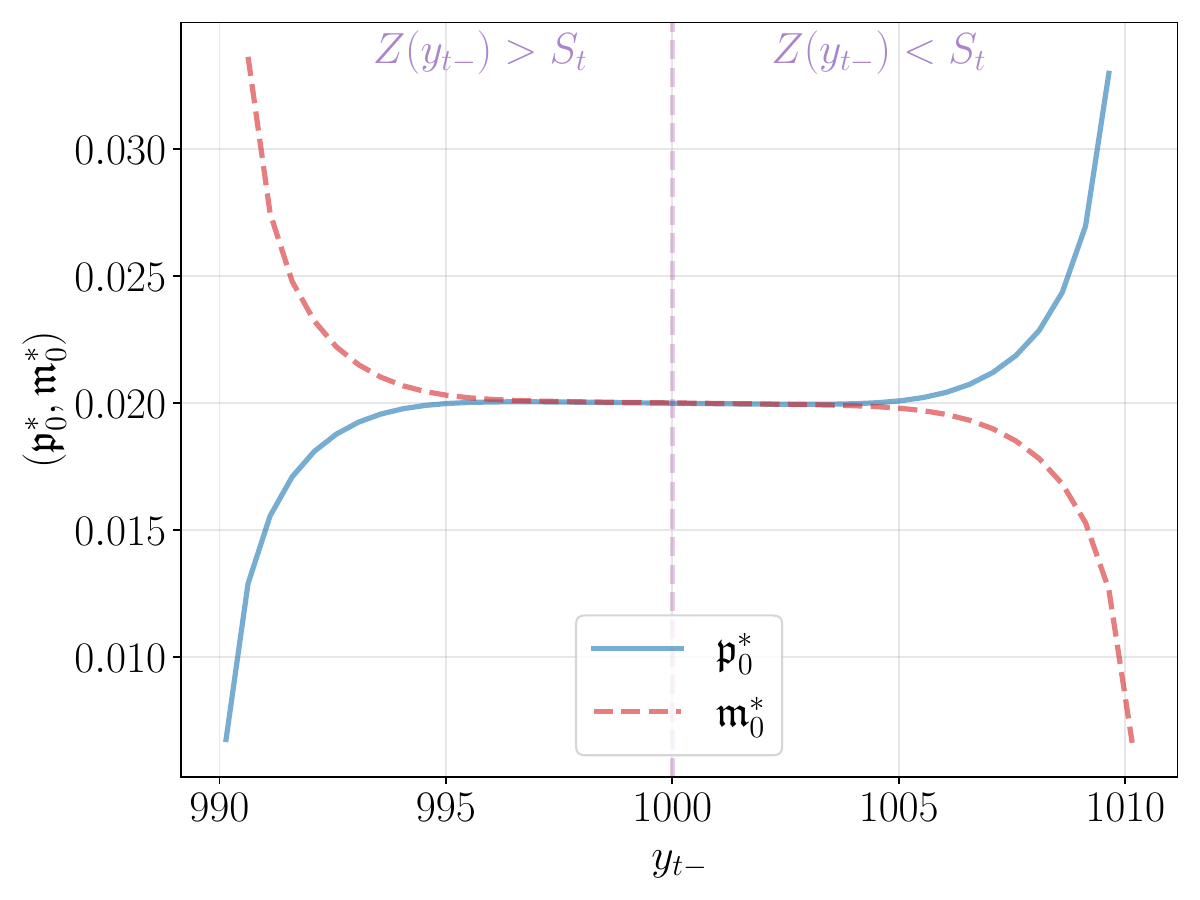}
        \caption{Optimal fees for selling $\feeLTsells_{0}^{*}(t,y_{t-})$ (solid line) and for buying $\feeLTbuys_{0}^{*}(t,y_{t-})$ (dashed line) at time $t=0.5$ as a function of the quantity of asset $Y$ in the pool.}
        \label{fig: optimal fees k=0}
    \end{subfigure}
\end{figure}

We see that when $k\to 0$ the optimal fee structure becomes constant in the central region. The only mechanism  that prevails is that of keeping the inventory away from the boundaries. Given that in the limit when $k\to 0$ the sensitivity of order arrivals to the centralised reference price disappears, one may interpret this scenario as one in which price formation happens in the venue.\footnote{Here one needs to account for scaling. Indeed what is being plotted in Figure \ref{fig: optimal fees k=0} is $k\,\feeLTsells$ and $k\,\feeLTbuys$, which accounts for the idea that as price formation happens in the AMM, arrivals will become sensitive to $\feeLTsells$ and $\feeLTbuys$ independently of the difference between the external price and the internal price. } Thus, we conclude that when price formation happens in the AMM the optimal fee is constant. Note that in practice one expects  the boundaries to be far from being reached.  

\begin{figure}[H]
    \centering
    \begin{subfigure}[b]{0.49\textwidth}
        \centering
        \includegraphics[width=\textwidth]{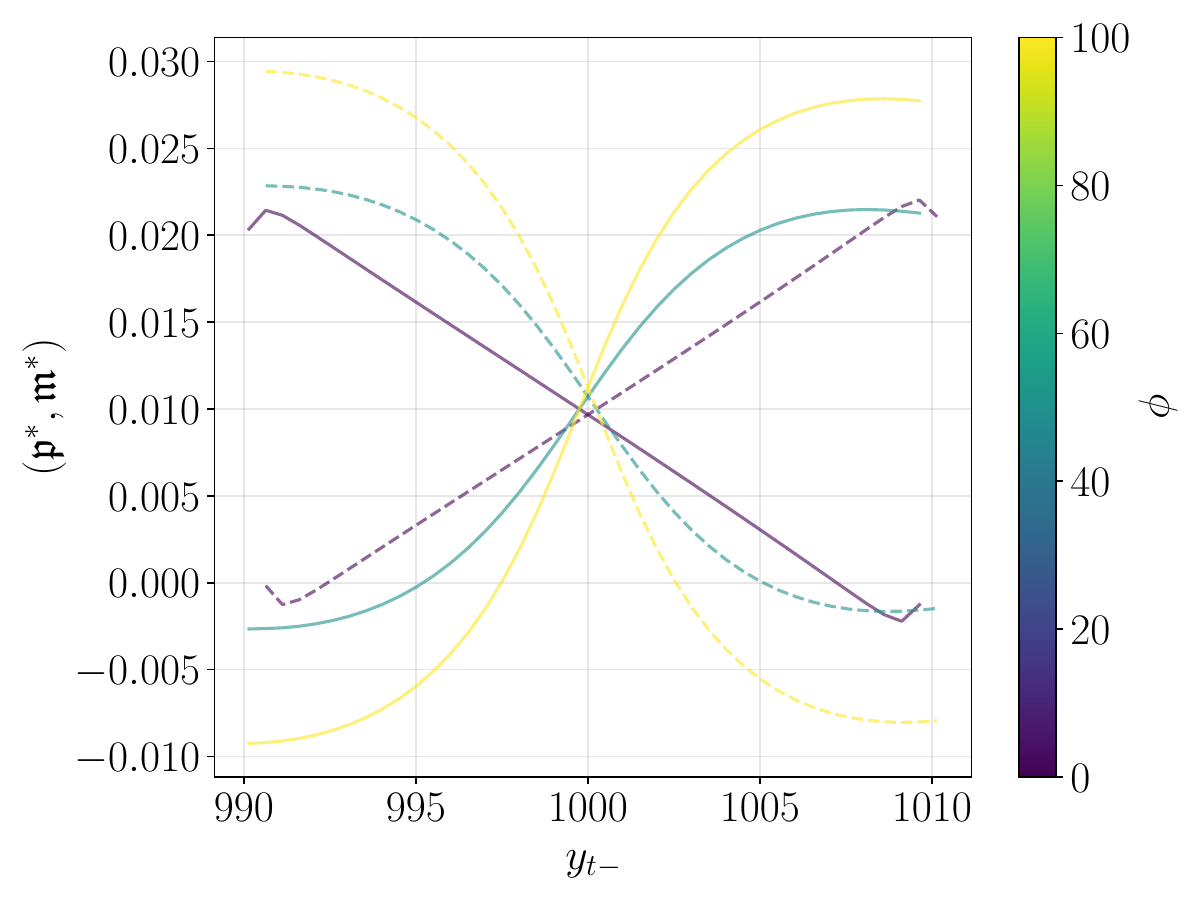}
        \caption{Optimal fees for selling $\feeLTsells^{*}(t,y_{t-},s)$ (solid line) and for buying $\feeLTbuys^{*}(t,y_{t-},s)$ (dashed line) at time $t=0.5$ and $s = 100$ as a function of the quantity of asset $Y$ in the pool and of $\pencons$ when $k = 2$.}
        \label{fig: optimal fees with penalty k=1}
    \end{subfigure}
    \hfill
    \begin{subfigure}[b]{0.49\textwidth}
        \centering
        \includegraphics[width=\textwidth]{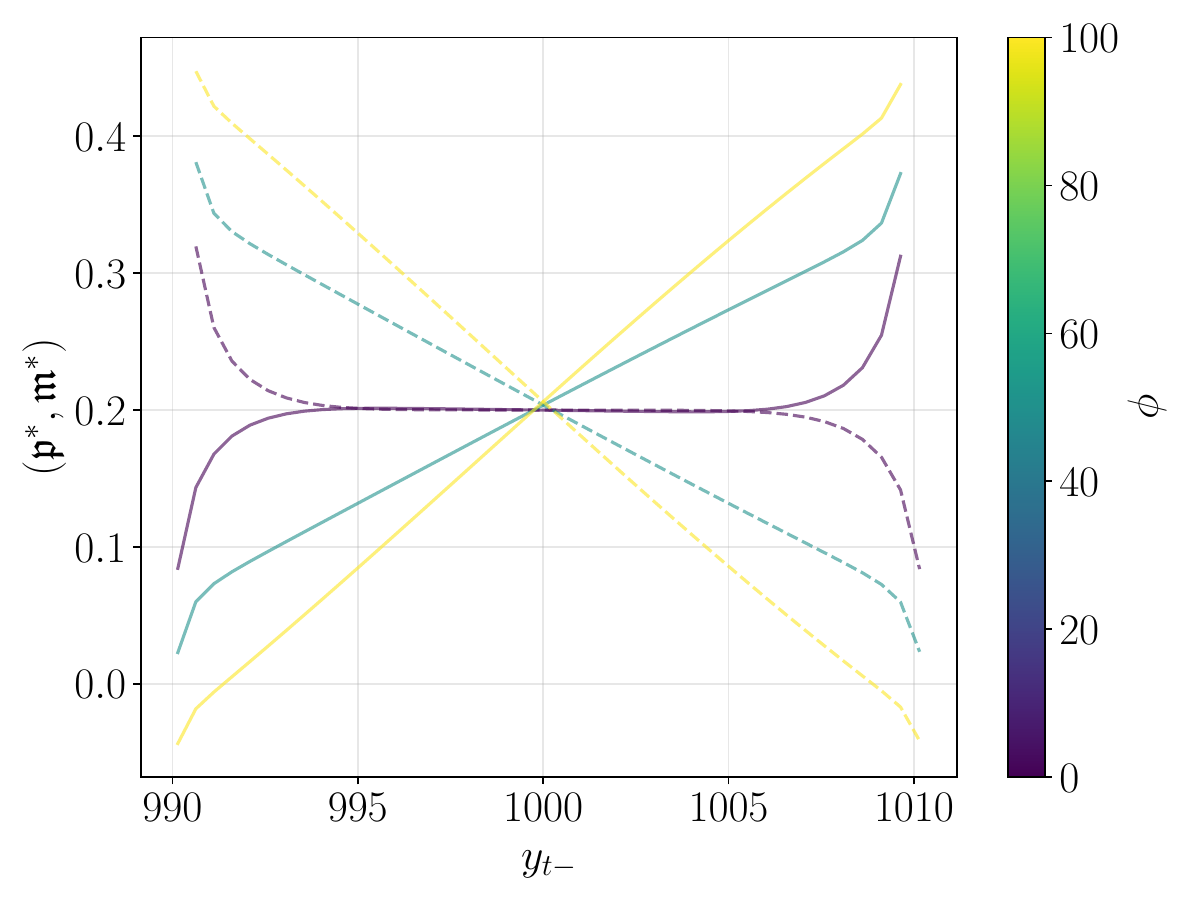}
        \caption{Optimal fees for selling $\feeLTsells^{*}(t,y_{t-},s)$ (solid line) and for buying $\feeLTbuys^{*}(t,y_{t-},s)$ (dashed line) at time $t=0.5$ and $s = 100$ as a function of the quantity of asset $Y$ and of $\pencons$ in the pool when $k = 0.1$.}
        \label{fig: optimal fees with penalty k=0.1}
    \end{subfigure}
\end{figure}

Next we discuss how the optimal fees behave as functions of $\pencons$. If we increase $\pencons$, the AMM tries to keep prices aligned to the external price by pushing quantities of asset $Y$ away from the boundaries. To this end, it charges low fees to sell and high fees to buy when the price is less than the centralised reference price and vice versa when the price is high. Figure \ref{fig: optimal fees with penalty k=1} and Figure \ref{fig: optimal fees with penalty k=0.1} show the fees as a function of quantity and penalization constant when $t=0.5$, for $k=2$ and $k=0.1$ respectively. The curves corresponding to $\pencons = 0$ are the ones in Figure \ref{fig: optimal fees} and Figure \ref{fig: optimal fees k=0.1} respectively .

\begin{figure}[H]
    \centering
    \begin{subfigure}[b]{0.48\textwidth}
        \centering
        \includegraphics[width=\textwidth]{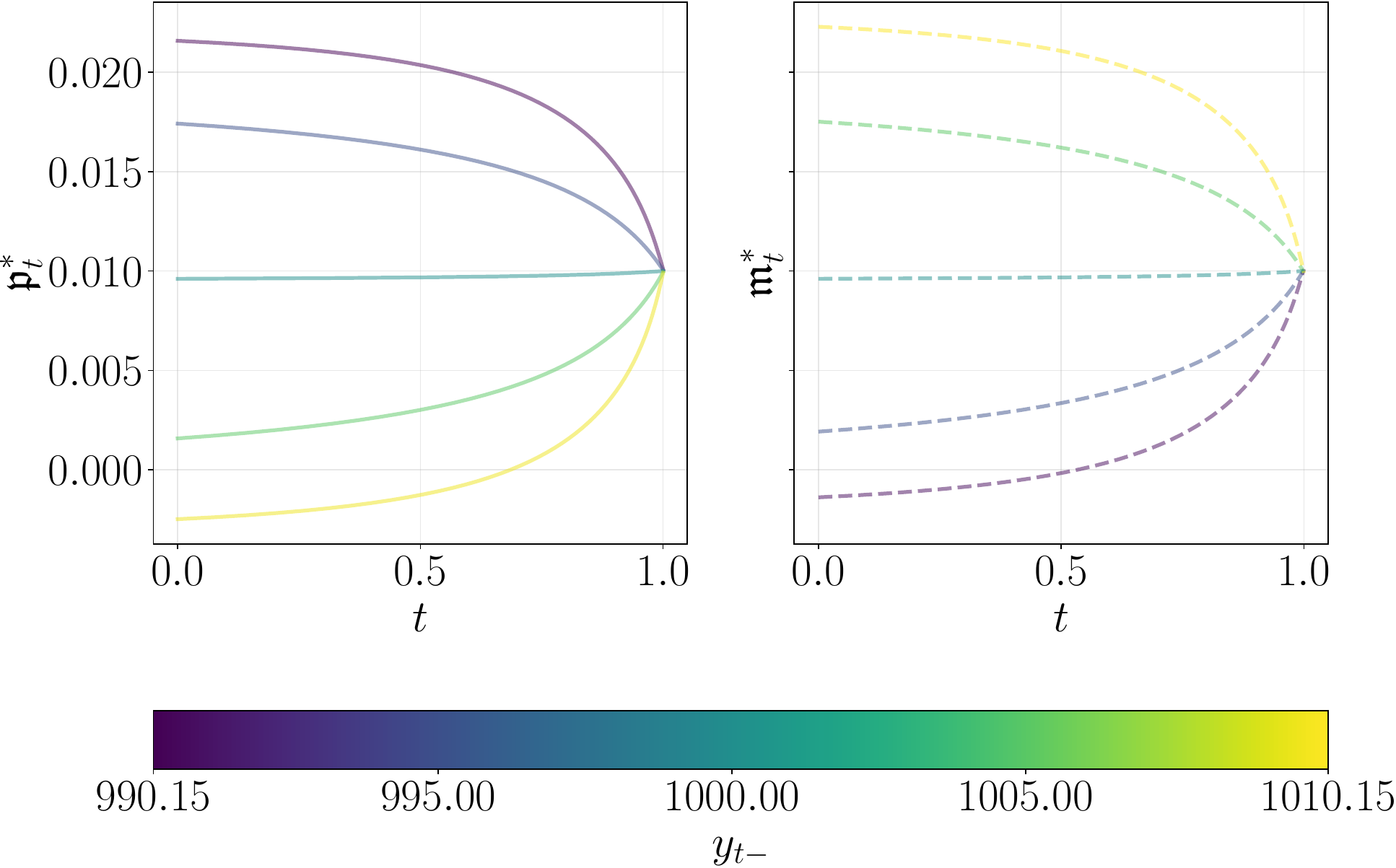}
        \caption{Optimal fees for selling $\feeLTsells^{*}(t,y_{t-},s)$ (solid line) and for buying $\feeLTbuys^{*}(t,y_{t-},s)$ (dashed line) at time $t=0.5$ and $s = 100$ as a function of the quantity of asset $Y$ (colorbar) in the pool and of time when $k = 2$.}
        \label{fig: optimal fees on time and quantity k=1}
    \end{subfigure}
    \hfill
    \begin{subfigure}[b]{0.48\textwidth}
        \centering
        \includegraphics[width=\textwidth]{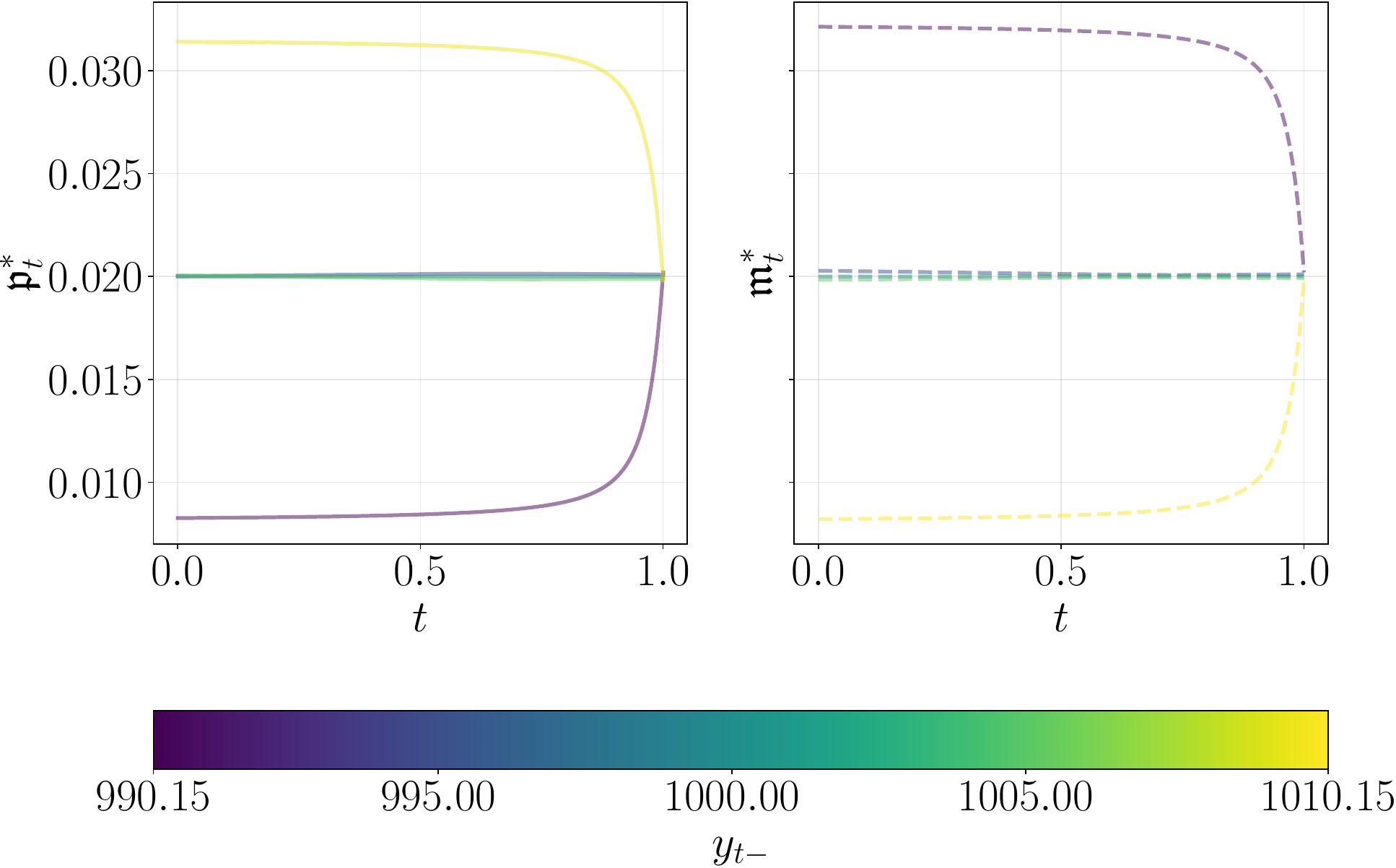}
        \caption{Optimal fees for selling $\feeLTsells^{*}(t,y_{t-},s)$ (solid line) and for buying $\feeLTbuys^{*}(t,y_{t-},s)$ (dashed line) at time $t=0.5$ and $s = 100$ as a function of the quantity of asset $Y$ (colorbar) and of time in the pool when $k = 0.1$.}
        \label{fig: optimal fees on time and quantity k=0.1}
    \end{subfigure}
\end{figure}

As the penalty parameter $\pencons$ increases, it alters both the shape and slope of the optimal fees. When $\pencons$ becomes sufficiently large, the penalty term dominates the allocation of the fees, with alignment of quotes becoming a secondary effect.
The following plots show the behaviour of the fees as a function of time ($x$ axis) and quantity in the pool (colorbar) for $k=2$ and $k=0.1$. The effect present in the Figure \ref{fig: optimal fees k=0.1} can be seen also in Figure \ref{fig: optimal fees on time and quantity k=0.1}.

Lastly we compare our optimal strategy with other different strategies:
\begin{itemize}
    \item[$(i)$] the linear strategy described in Figure \ref{fig: linear fees}, where at each time with linearised $\feeLTsells^{\mathrm{lin}}(t,y,s)$, $\feeLTbuys^{\mathrm{lin}}(t,y,s)$ in a neighbourhood of $y_0$, and
    \item[$(ii)$] the \emph{constant} strategy where the fees are constant for every time $t$ and every quantity $y$. The constant $c$ is chosen as the average of the optimal two fees at $t=0.5$ and $s = 100$ for $y = y_0$, i.e., \[ c = \frac{\feeLTsells^{*}(0.5,y_0,s_0) + \feeLTbuys^{*}(0.5,y_0,s_0)}{2}. \]
\end{itemize}

We consider the following parameters $\depth = 10^{8}$, $y_0 = 1000$, $s_0 = 100$, $\pencons = 0$, $\underline{y} = 980$, $\overline{y} = 1020$, $T=1$, $\sigma = 1$ and we run simulations for different values of $\intbuy$, $\intsell$ and $k$. We carry out 100,000 simulations and we discretise $[0,T]$ in $1,000$ timesteps.\footnote{Our code is publicly available at \url{https://github.com/leonardobaggiani/amm-fees.git}.} 

We observe that the revenues are decreasing in the values of $\expdecay$ and increasing in the value of $\intsell$ and $\intbuy$. The latter is because as $\intsell$ and $\intbuy$ increase the number of orders increase and hence the revenue increases, similarly, as $\expdecay$ decreases, the arbitrageurs react less to mispricing and simultaneously, the optimal fees increase and thus the AMM collects more fees.

In the following table, the column \textit{fees} shows the revenue from collecting fees, the column \textit{sell} shows the number of sell orders, the column \textit{buy} shows the number of buy orders, and the column \textit{QV} shows the quadratic variation of the instantaneous exchange rate $Z$ defined in \eqref{eq: marginal exchange}.
 
\begin{center}
\begin{tabular}[h]{@{}rcccccccc@{}}
\toprule
& \multicolumn{4}{c}{$\lambda^{+} = \lambda^{-} = 100$}
& \multicolumn{4}{c}{$\lambda^{+} = \lambda^{-} = 150$} \\
\cmidrule(lr){2-5} \cmidrule(lr){6-9}
& fees & sell & buy & QV
& fees & sell & buy & QV\\
\midrule

\multicolumn{9}{l}{$k=2$}\\
\midrule
Optimal   & 35.61 & 36.05 & 35.89 & 0.69 & 53.00 & 53.59 & 53.40 & 1.01 \\
Linear    & 35.61 & 36.04 & 35.89 & 0.69 & 53.00 & 53.58 & 53.42 & 1.01 \\
Constant  & 35.21 & 35.23 & 35.19 & 0.68 & 52.31 & 52.32 & 52.29 & 0.99 \\

\midrule
\multicolumn{9}{l}{$k=1$}\\
\midrule
Optimal   & 71.59 & 36.06 & 35.94 & 0.69 & 106.47 & 53.59 & 53.46 & 1.01 \\
Linear    & 71.59 & 36.05 & 35.95 & 0.69 & 106.47 & 53.58 & 53.48 & 1.01 \\
Constant  & 71.32 & 35.69 & 35.64 & 0.69 & 105.99 & 53.01 & 52.98 & 1.00 \\
\bottomrule
\end{tabular}
\end{center}

We find that optimal strategy outperforms the constant strategy and the linear approximation of the fees has a performance that is indistinguishable from that of the optimal strategy for the above parameters.
Thus, a linear fee structure is suitable when designing the fees to charge in these venues.

When $\pencons = 0$, the fee mechanism in our model does not directly force the instantaneous marginal price of the pool to track the centralised reference price $S_t$. Indeed, the fees only affect the effective exchange rates faced by liquidity takers, while the marginal price $Z_t$ is entirely determined by the AMM curve. As a consequence, whenever the centralised reference price lies inside the fee-adjusted spread, there is no arbitrage-induced increase in the intensities forcing the pool inventory to move, and hence no reason for $Z_t$ to remain close to $S_t$.

\begin{figure}[H]
    \centering
    \includegraphics[width=0.6\textwidth]{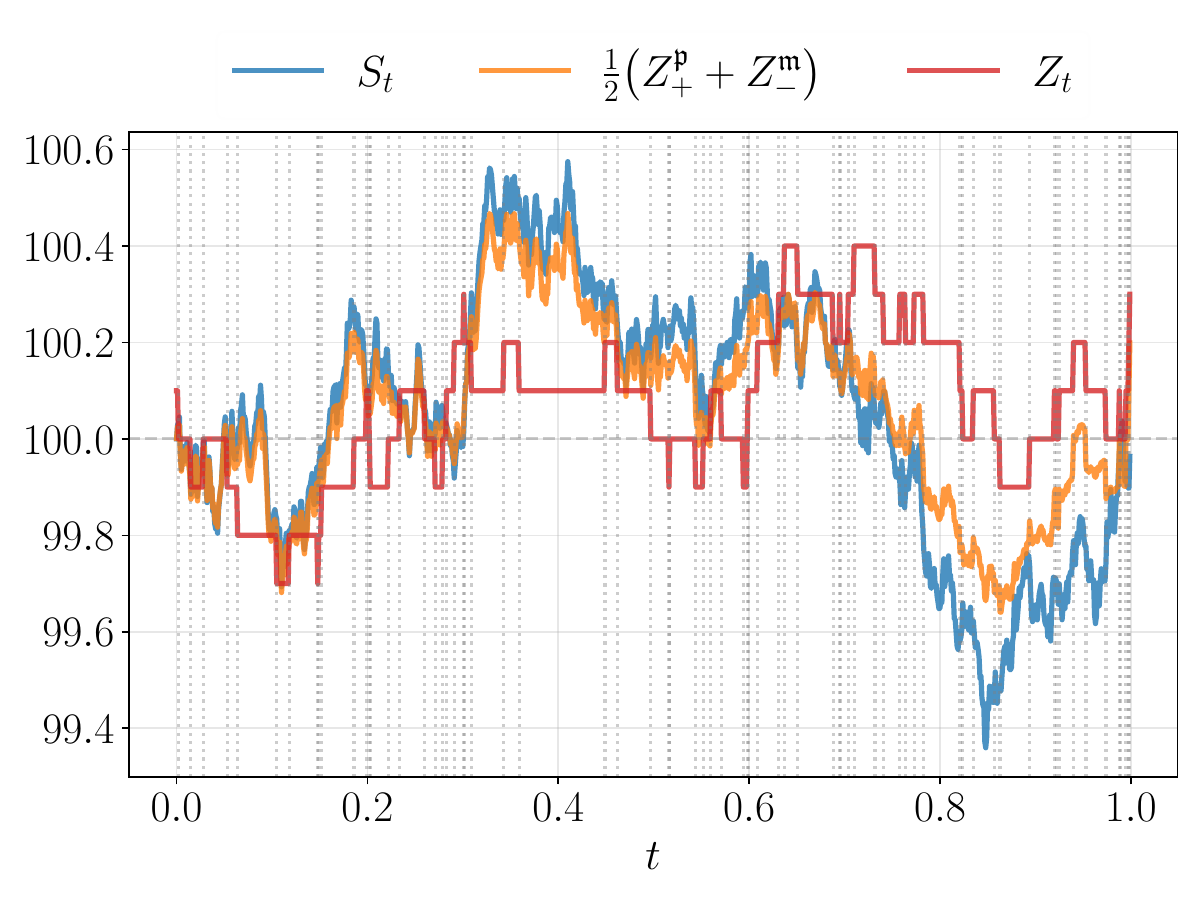}
    \caption{centralised reference price $S_t$ and marginal pool price $Z_t$ when $\pencons=0$.}
    \label{fig:no-penalty-tracking}
\end{figure}

What happens instead is that $S_t$ remains close to the midpoint between the fee-adjusted buy and sell rates. This can be observed in Figure~\ref{fig:no-penalty-tracking}. We can observe that the marginal pool price $Z_t$ does not track the centralised reference price particularly closely. In contrast, the figure shows that the midpoint of the fee-adjusted spread tracks the centralised reference price much more accurately.

\begin{figure}[H]
    \centering
    \includegraphics[width=0.6\textwidth]{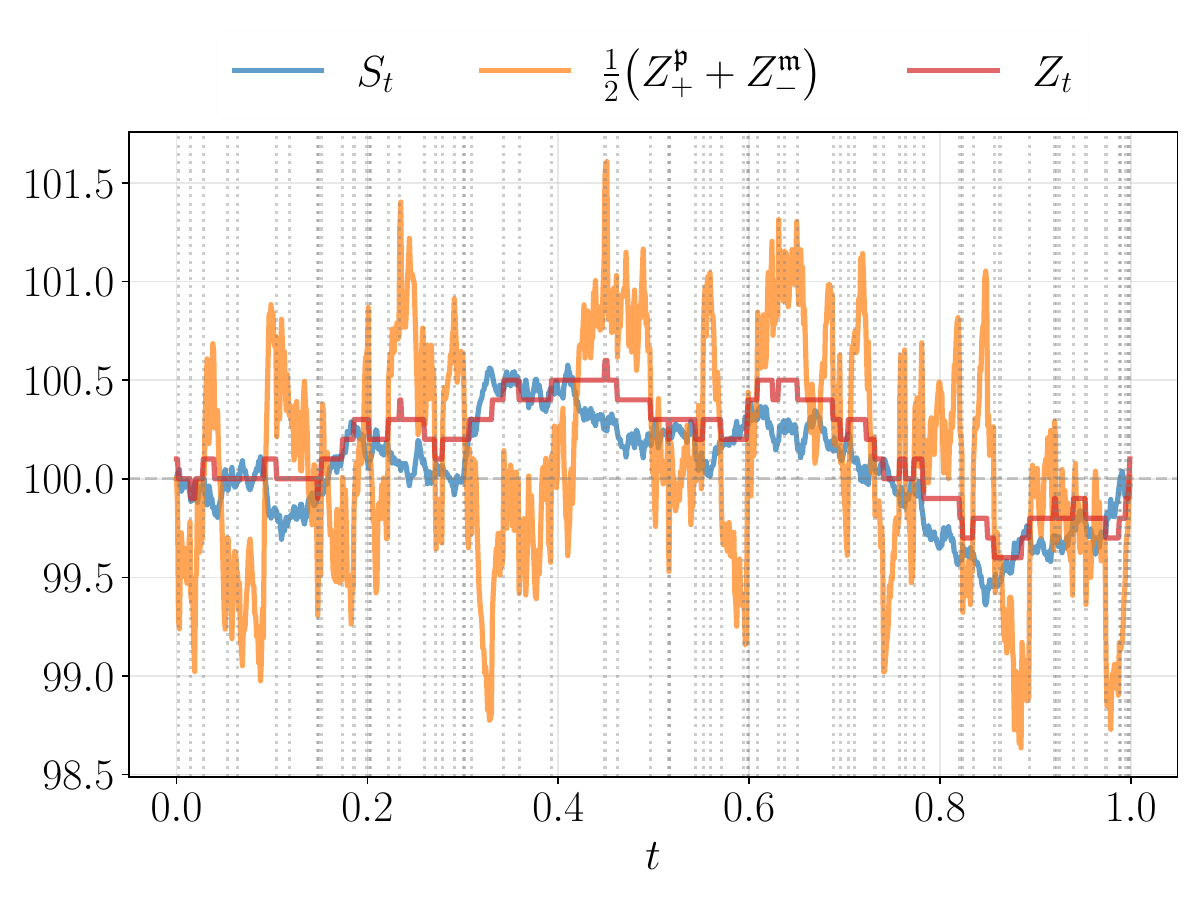}
    \caption{Centralised reference price $S_t$ and marginal pool price $Z_t$ when $\pencons>0$. With penalization, the marginal pool price tracks the centralised reference price more closely.}
    \label{fig:penalty-tracking}
\end{figure}

If one wants the marginal price inside the pool to track the centralised reference price, it is sufficient to choose a non-zero penalization parameter $\pencons$ in the running penalty
\[
    \Pi(Y_t^{\feeLTsells,\feeLTbuys},S_t)
    =
    \pencons\bigl(Z(Y_t^{\feeLTsells,\feeLTbuys})-S_t\bigr)^2.
\]
This term penalizes deviations between the marginal pool price and the centralised reference price, incentivizing the AMM to choose fees that push the inventory toward the oracle-implied state. As a consequence, the marginal price $Z_t$ now follows the centralised reference price much more closely at the expense of a rougher version of $\frac{1}{2} \left( \exratebuy^{\feeLTbuys} + \exratesell^{\feeLTsells} \right) $. This effect is illustrated in Figure~\ref{fig:penalty-tracking}.

The drawback is that enforcing this alignment might create additional arbitrage opportunities. Indeed, keeping the marginal pool price close to the centralised reference price requires more aggressive fee distortions, which generate larger fluctuations in the fee-adjusted buy and sell rates around the centralised reference price. 

\section{Quadratic approximation of the order arrival rates}\label{section: second Approximation}

For the second approximate solution we return to the original HJB equation in \eqref{eq: eq for g}. 
Here, we make a quadratic approximation of the exponential terms:
\small{
\begin{align*}
    e^{\expdecay (-\oracleprice \deltasell(y) + \exratesell (y) \deltasell(y) + g(t,y + \deltasell(y),s) - g(t,y,s) )} \approx 1 & + \expdecay (-\oracleprice \deltasell(y) + \exratesell (y) \deltasell(y) + g(t,y + \deltasell(y) ,s) - g(t,y,s) ) \\
    & + \frac{1}{2} [ \expdecay (-\oracleprice \deltasell(y)  + \exratesell (y) \deltasell(y) + g(t,y + \deltasell(y),s) - g(t,y,s) ) ]^{2}, \\
    e^{ \expdecay( \oracleprice \deltabuy(y) - \exratebuy (y) \deltabuy(y)   + g(t,y - \deltabuy(y) ,s) - g(t,y,s))} \approx 1 & + \expdecay( \oracleprice \deltabuy(y)  - \exratebuy (y) \deltabuy(y)  + g(t,y - \deltabuy(y),s) - g(t,y,s)) \\
    & + \frac{1}{2} [\expdecay( \oracleprice \deltabuy(y) - \exratebuy (y) \deltabuy(y) + g(t,y - \deltabuy(y),s) - g(t,y,s))]^{2}.
\end{align*}
}
The HJB in \eqref{eq: eq for g} becomes 

\begin{equation}\label{eq: approx eq for g}
\begin{aligned}
    \frac{\partial}{\partial t}  g(t,y,s) & + \frac{\sigma^{2}}{2} \frac{\partial^{2}}{\partial s^{2}} g(t,y,s) - \pencons(Z(y) -s)^{2} + \\
    & \frac{e^{-1} \intsell}{\expdecay} [ 1 - \expdecay(\oracleprice \deltasell(y) + \exratesell(y) \deltasell(y) + g(t,y + \deltasell(y),s) - g(t,y,s)) \\
    & + \frac{\expdecay^{2}}{2}(- \oracleprice \deltasell(y) + \exratesell(y) \deltasell(y) + g(t,y + \deltasell(y),s) - g(t,y,s))^{2} ] \\
    & \frac{e^{-1} \intbuy}{\expdecay} [ 1 + \expdecay(\oracleprice \deltabuy(y) - \exratebuy(y) \deltabuy(y) + g(t,y - \deltabuy(y),s) - g(t,y,s)) \\
    & + \frac{\expdecay^{2}}{2}( \oracleprice \deltabuy(y) - \exratebuy(y)\deltabuy(y)  + g(t,y - \deltabuy(y),s) - g(t,y,s))^{2} ] = 0.
\end{aligned}
\end{equation}

We assume that $\deltabuy(y)$ and $\deltasell(y)$ are constant equal to $\delta^{-}$ and $\delta^{+}$, respectively.

\begin{remark}
    The assumption that $\deltabuy(y)$ and $\deltasell(y)$ are constant is necessary if one wishes to carry out a quadratic expansion of the exponential terms. If we employ a linear approximation of the exponential terms, we could work with the case where $\deltabuy(y)$ and $\deltasell(y)$ are linear in $y$.
\end{remark}

Lastly, we use a linear approximation to the exchange rates

\begin{align*}
    Z(y) & = \frac{\depth}{y^{2}} \approx \frac{\depth}{y_0^2} - 2(y - y_0) \frac{\depth}{y_0^3}, \\
    \exratesell(y) \delta^{+}& =  \frac{\depth}{y^{2} + \delta^{+} y} \approx \frac{\depth}{y_0^2 + \delta^{+} y_0} - (y - y_0) \frac{\depth(2 y_0 + \delta^{+})}{(y_0^2 + \delta^{+}y_0)^2}, \\
    \exratebuy(y) \delta^{-} & =  \frac{\depth}{y^{2} - \delta^{-}y} \approx \frac{\depth}{y_0^2 - \delta^{-}y_0} - (y - y_0) \frac{\depth(2 y_0 - \delta^{-})}{(y_0^2 - \delta^{-}y_0)^2}.
\end{align*}

In order to find a solution for the equation \eqref{eq: approx eq for g} we assume that $g(t,y,s)$ is a quadratic polynomial in $y$ of the form \[ g(t,y,s) = y^{2} A(t) + y B(t,s) + C(t,s), \]
with $A(T) = B(T,s) = C(T,s) = 0$. Proposition \eqref{thm: solution to the PDE system} formalises the solution to the HJB equation.

\begin{theorem}\label{thm: solution to the PDE system}
    There exists constants\footnote{The full definition of the constants $\{ \psi_{i} \}_{1 \leq i \leq 26}$ is publicly available in the GitHub of the project.} $\{ \psi_{i} \}_{1 \leq i \leq 26} \subset \mathbb{R}$ that define the unique solution to the HJB equation \eqref{eq: approx eq for g} as follows. Let $A:[0,T] \to \mathbb{R}$ be the solution to the Riccati ODE
    \begin{equation*}
        \psi_0 + \psi_1 A(t) + \psi_2 A^{2}(t) + A'(t), \quad \quad A(T) = 0.
    \end{equation*}
    Let $b_0$, $b_1 : [0,T] \to \mathbb{R}$ be the unique solutions to the system
    \[
    \begin{cases}
       \psi_3 + \psi_4 A(t) + \psi_5 A^{2}(t) + \psi_6 A(t) b_0(t) + \psi_7 b_0(t) + b_0^{'}(t) = 0 \\
       \psi_8 + \psi_9 A(t) + \psi_{10}A(t)b_1(t) + \psi_{11}b_1(t) + b_1^{'}(t)= 0,
    \end{cases}
    \]
    with terminal conditions $b_0(T) = b_1(T) = 0$.
    Let $c_0$, $c_1$, $c_2: [0,T] \to \mathbb{R}$ be the unique solutions to the system
    \[
    \begin{cases}
        \psi_{12} + \psi_{13} A(t) + \psi_{14}A^{2}(t) + \psi_{15}A(t)b_0(t) + \psi_{16}b_0(t) + \psi_{17}b_0^{2}(t) + \sigma^{2} c_{2}(t) + c_0^{'}(t) = 0, \\
        \psi_{18} + \psi_{19}A(t) + \psi_{20}A(t)b_{1}(t) + \psi_{21}b_{0}(t) + \psi_{22}b_0(t)b_{1}(t) + \psi_{23}b_{1}(t) + c_{1}^{'}(t) = 0, \\
        \psi_{24} + \psi_{25} b_{1}(t) + \psi_{26}b_{1}^{2}(t) + c_{2}^{'}(t) = 0,
    \end{cases}
    \]
    with terminal conditions $c_0(T) = c_1(T) = c_2(T) = 0$.
    Define $\hat{B},\hat{C} : [0,T] \times \mathbb{R} \to \mathbb{R}$ as
    \begin{align*}
        \hat{B}(t,s) & = s b_{1}(t) + b_0(t), \\
        \hat{C}(t,s) & = s^{2} c_{2}(t) + s c_1(t) + c_0(t).
    \end{align*}
    Then, the function \[ g(t,y,s) = y^{2} \hat{A}(t) + y \hat{B}(t,s) + \hat{C}(t,s), \] is a solution to the HJB equation \eqref{eq: approx eq for g}.
\end{theorem}

With the above results, the optimal fees from \eqref{eq: maximizers} become

\begin{equation}\label{eq: maximizers second approx}
\begin{aligned}
    \feeLTsells^{*}(t,y) & := -\frac{(2y + \delta^{+})\hat{A}(t) + \hat{B}(t,s)}{ \exratesell(y) } + \frac{1}{\expdecay \exratesell(y) \delta^{+}}, \\
    \feeLTbuys^{*}(t,y) & := -\frac{(-2y + \delta^{-})\hat{A}(t) - \hat{B}(t,s)}{ \exratebuy(y)} + \frac{1}{\expdecay \exratebuy(y) \delta^{-}}.
\end{aligned}
\end{equation}

Observe that the fees depend linearly on $y$ and $s$ and they do not depend on $\sigma$.

\subsection{Optimal fee structure: dynamic centralised reference price with second order approximation}\label{section: Optimal fee structure: dynamic external price with second order approximation}

We perform numerical simulations of the results found in Theorem \ref{thm: solution to the PDE system}. In order to make it consistent with the hypothesis of the theorem, we assume that the traded amounts are $\deltasell(y) = \delta^{+} = 0.5$ and $\deltabuy(y) = \delta^{-} = 0.5$. We fix an initial value of $y_0 = 1000$ which gives the following grid for $y$ as \[ y^{i} = (1000 + i), \quad  \quad \text{ for } i \in \{-20, -19.5, \dots, 19.5, 20 \}. \]
We take the time horizon $T=1$, the two baseline intensities $\intbuy = \intsell = 50$, the rate of exponential decay $\expdecay = 2$, the centralised reference price $S_{t} = S_0 + \sigma W_{t}$ with $S_0 = 100$, $\sigma = 0.2$ and $\pencons = 0$. As before, we assume that the depth of the pool is $\depth = 10^{8}$.

The next figures show the plot for the optimal fees of \eqref{eq: maximizers second approx} and the comparison with the optimal fees obtained with the first approximation. It is important to point out that the fees in \eqref{eq: maximizers second approx} are no longer skewed at the boundaries.

\begin{figure}[H]
    \centering
    \begin{subfigure}[b]{0.49\textwidth}
        \centering
        \includegraphics[width=\textwidth]{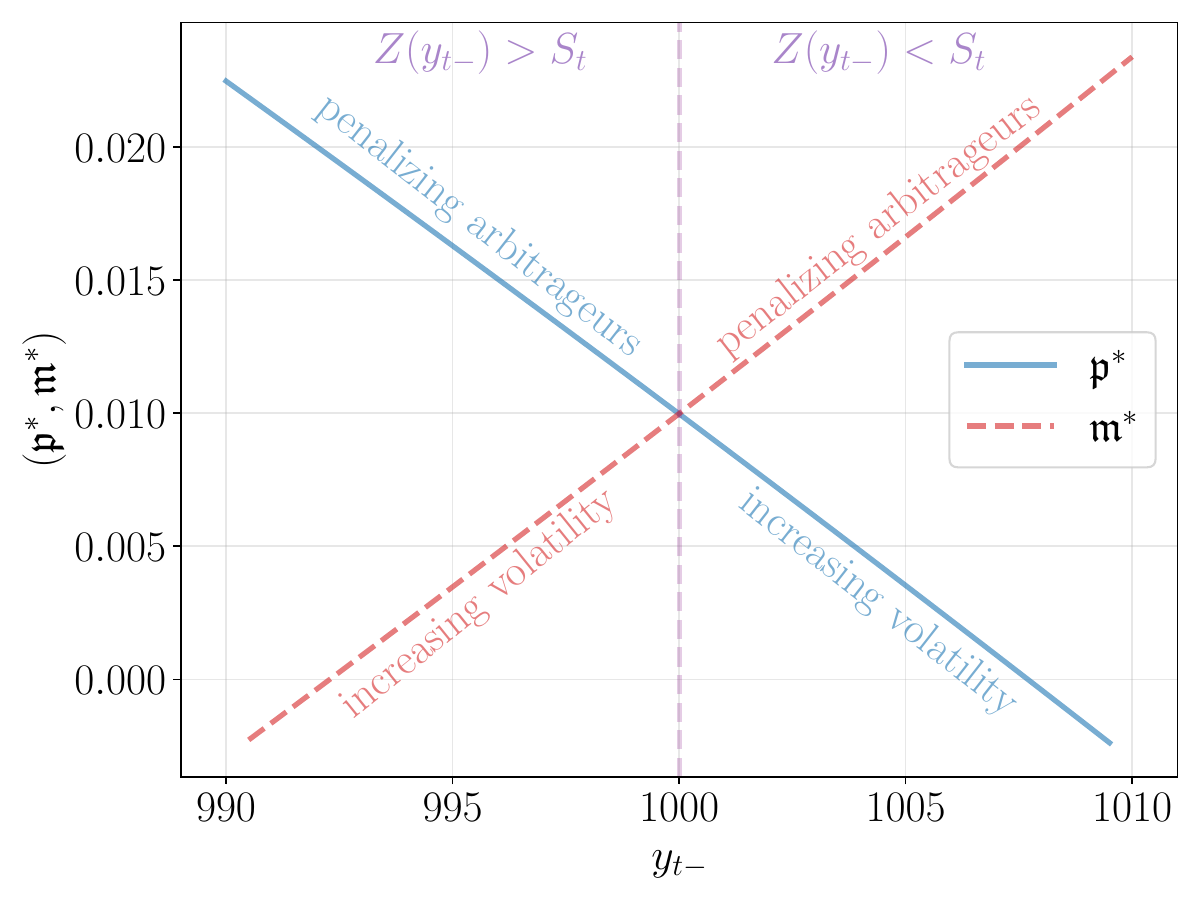}
        \caption{Optimal fees (second approximation Section \ref{section: second Approximation}) for selling $\feeLTsells^{*}(t,y_{t-})$ (solid line) and for buying $\feeLTbuys^{*}(t,y_{t-})$ (dashed line) at time $t=0.5$ as a function of the quantity of asset $Y$ in the pool.}
        \label{fig: optimal fees second approx}
    \end{subfigure}
    \hfill
    \begin{subfigure}[b]{0.49\textwidth}
        \centering
        \includegraphics[width=\textwidth]{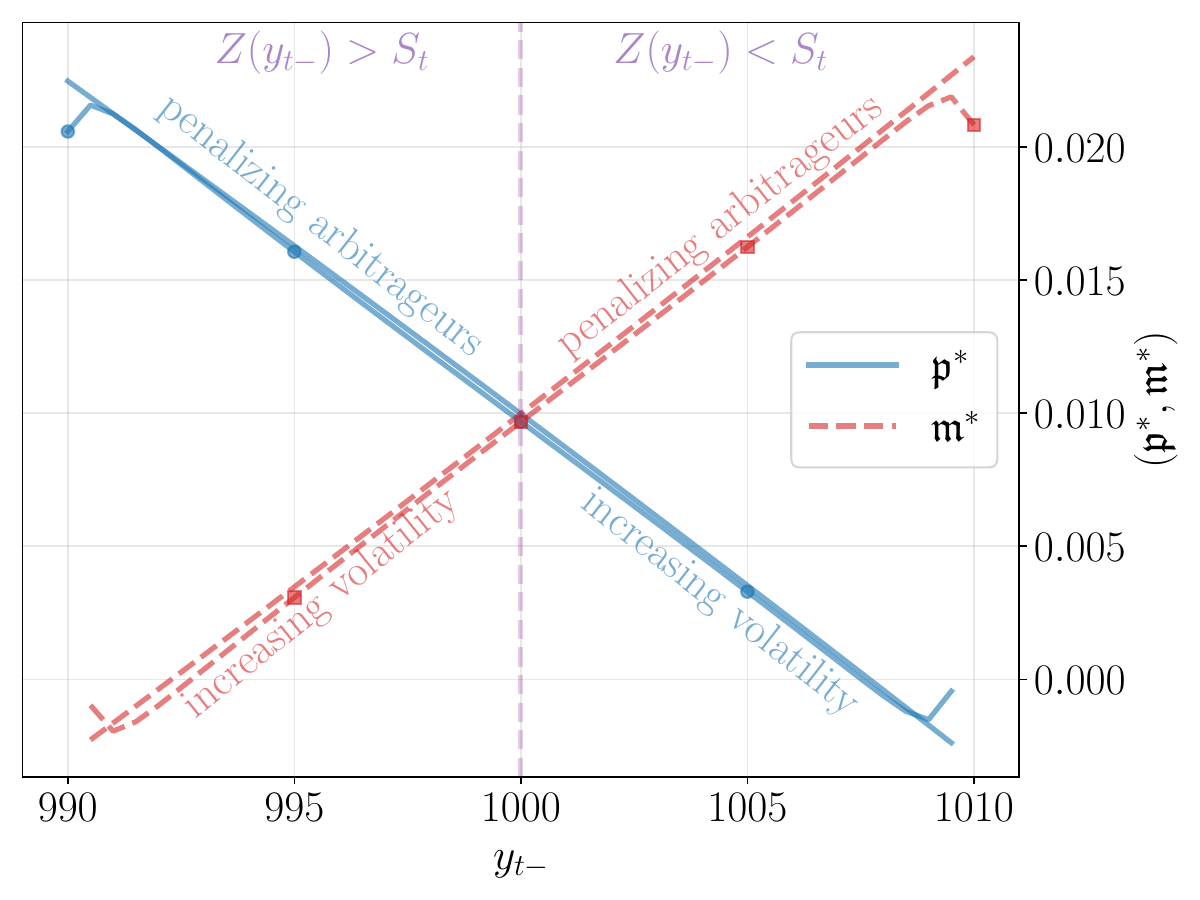}
        \caption{Optimal fees (first approximation Section \ref{section: First Approximation}) for selling $\feeLTsells^{*}(t,y_{t-})$ (solid line) and for buying $\feeLTbuys^{*}(t,y_{t-})$ (dashed line) at time $t=0.5$ as a function of the quantity of asset $Y$ in the pool.}
        \label{fig: optimal fees second approx vs optimal fees first approx}
    \end{subfigure}
\end{figure}

The plot in Figure \ref{fig: optimal fees second approx} confirm the insights we obtained from Figure \ref{fig: optimal fees}, namely that there are two distinct regimes for the fee structure: one in which arbitrageurs are penalised and one that aim to increase volatility and noise trading.
Next, we study the case when $k \to 0 $. Here, the second order approximation does not work well for smaller values of the parameters $\expdecay$. Indeed, since the fees are independent of $\sigma$ we would expect that as $k \to 0$ the fees in \eqref{eq: maximizers second approx} should resemble the one in Figure \ref{fig: optimal fees k=0}, however, the next two figures show that this is not the case.
The explanation for this phenomenon comes from the fact that when $k \to 0$ then $A(t) \to 0$ and $B(t) \to \psi_{3}t$, hence the following holds.

\begin{corollary}\label{cor: fee for k to 0 second approximation}
    Let $\feeLTsells^{*}(t,y)$ and $\feeLTbuys^{*}(t,y)$ be the optimal fees from Theorem \ref{eq: maximizers second approx} and define the quantities
    \begin{align}
    \feeLTsells^{*}_0(t,y^i) & : =  \frac{1}{ \exratesell(y^{i}) \delta^{+}}  \\
    \feeLTbuys^{*}_0(t,y^i) & : = \frac{1}{ \exratebuy(y^{i}) \delta^{-}} .
\end{align}
Then we have that 
\begin{align}
    \lim_{k \to 0} k\feeLTsells^{*}(t,y^i) = \feeLTsells^{*}_0(t,y^i) \quad \quad \text{ and } \quad \quad  \lim_{k \to 0} k\feeLTbuys^{*}(t,y^i) = \feeLTbuys^{*}_0(t,y^i),
\end{align}
for every $t \in [0,T]$ and $y^{i} \in \{ y^{-N}, \dots, y^{N} \}$.
\end{corollary}

\begin{figure}[H]
    \centering
    \begin{subfigure}[t]{0.49\textwidth}
        \centering
        \includegraphics[width=\textwidth]{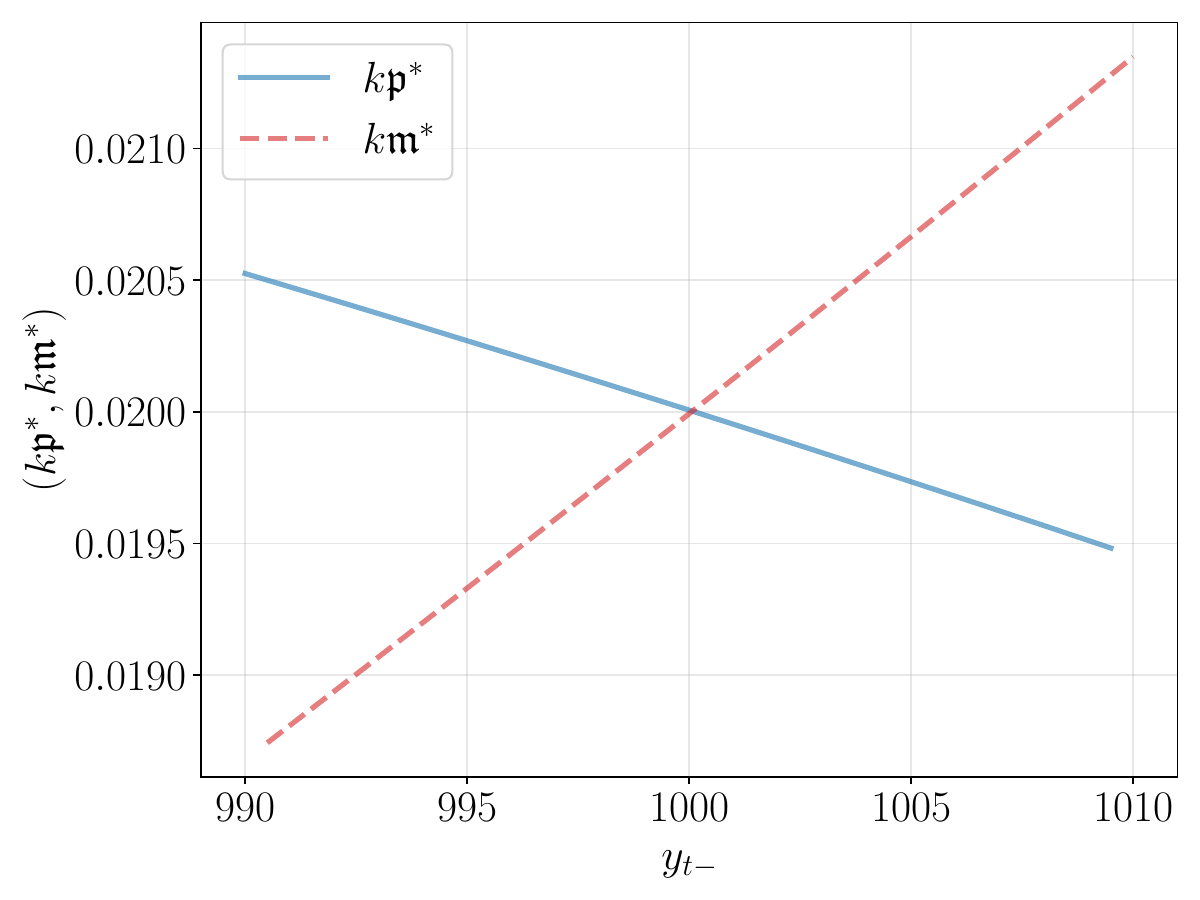}
        \caption{Optimal fees for selling $\feeLTsells^{*}(t,y_{t-})$ (solid line) and for buying $\feeLTbuys^{*}(t,y_{t-})$ (dashed line) for $k=0.25$ at time $t=0.5$ as a function of the quantity of asset $Y$ in the pool.}
        \label{fig: optimal fees second approx k=0.25}
    \end{subfigure}
    \hfill
    \begin{subfigure}[t]{0.49\textwidth}
        \centering
        \includegraphics[width=\textwidth]{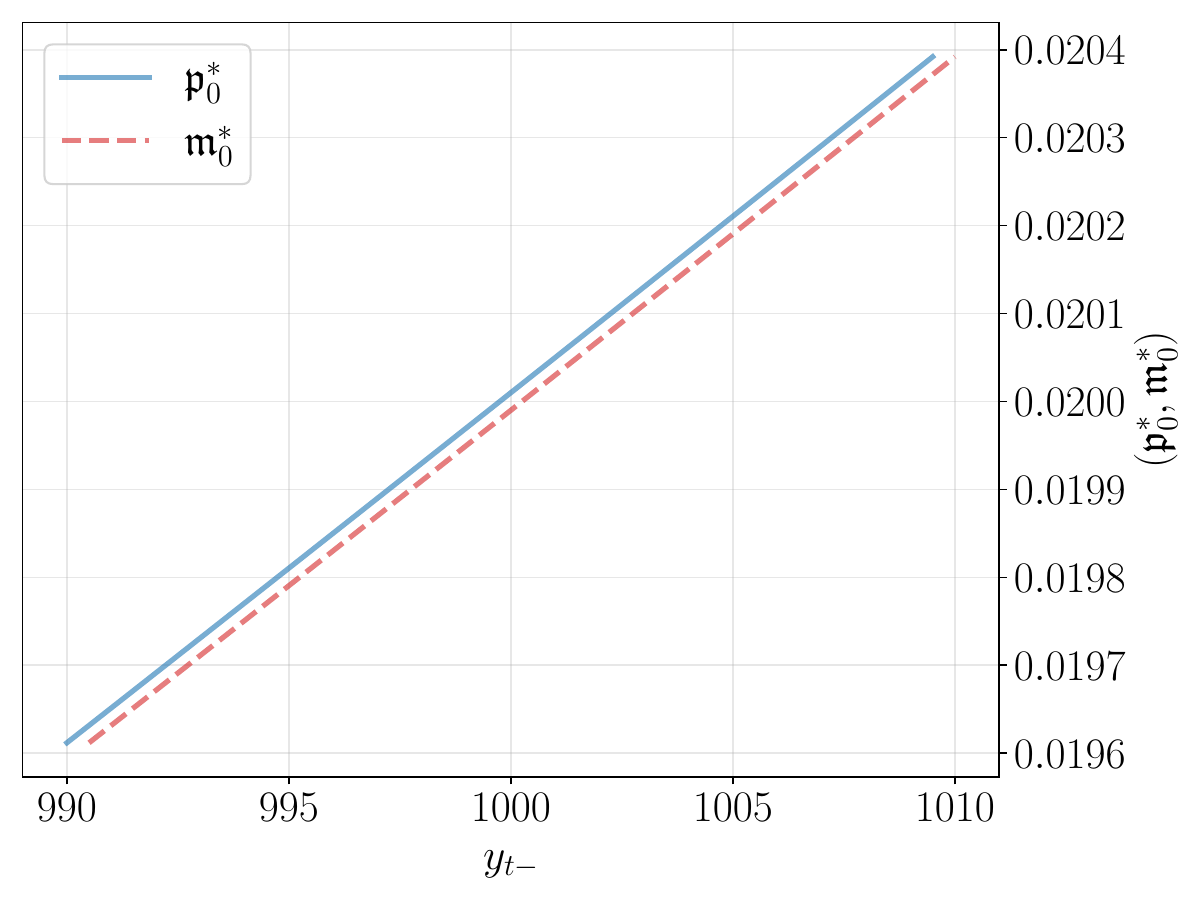}
        \caption{Optimal fees for selling $\feeLTsells_{0}^{*}(t,y_{t-})$ (solid line) and for buying $\feeLTbuys_{0}^{*}(t,y_{t-})$ (dashed line) at time $t=0.5$ as a function of the quantity of asset $Y$ in the pool.}
        \label{fig: optimal fees second approx k=0}
    \end{subfigure}
\end{figure}

Figures \ref{fig: optimal fees second approx k=0.25} and \ref{fig: optimal fees second approx k=0} illustrate the limiting behaviour described in Corollary \eqref{cor: fee for k to 0 second approximation}.

The next two figures show how the optimal fees behave as functions of $\pencons$ (left) and as a function of $S_t$ (right). Similarly to the first approximation there are two different regimes, one for $\pencons=0$ and one for $\pencons \neq 0$. Indeed, when $\pencons \neq 0$ the AMM tries to push the price away from the boundaries. Hence, when the quantity of asset $y$ is close to $\underline{y}$ it will charge high fees to buy and low (or even negative) fees to sell and when the quantity of asset $y$ is close to $\overline{y}$ we get the symmetric effect. As expected from the expression in \eqref{eq: maximizers second approx}, the dependence on $S_t$ is linear, furthermore, $\feeLTsells$ decreases when $S_{t}$ increases and $\feeLTbuys$ decreases as $S_{t}$ increases.

\begin{figure}[H]
    \centering
    \begin{subfigure}[t]{0.49\textwidth}
        \centering
        \includegraphics[width=\textwidth]{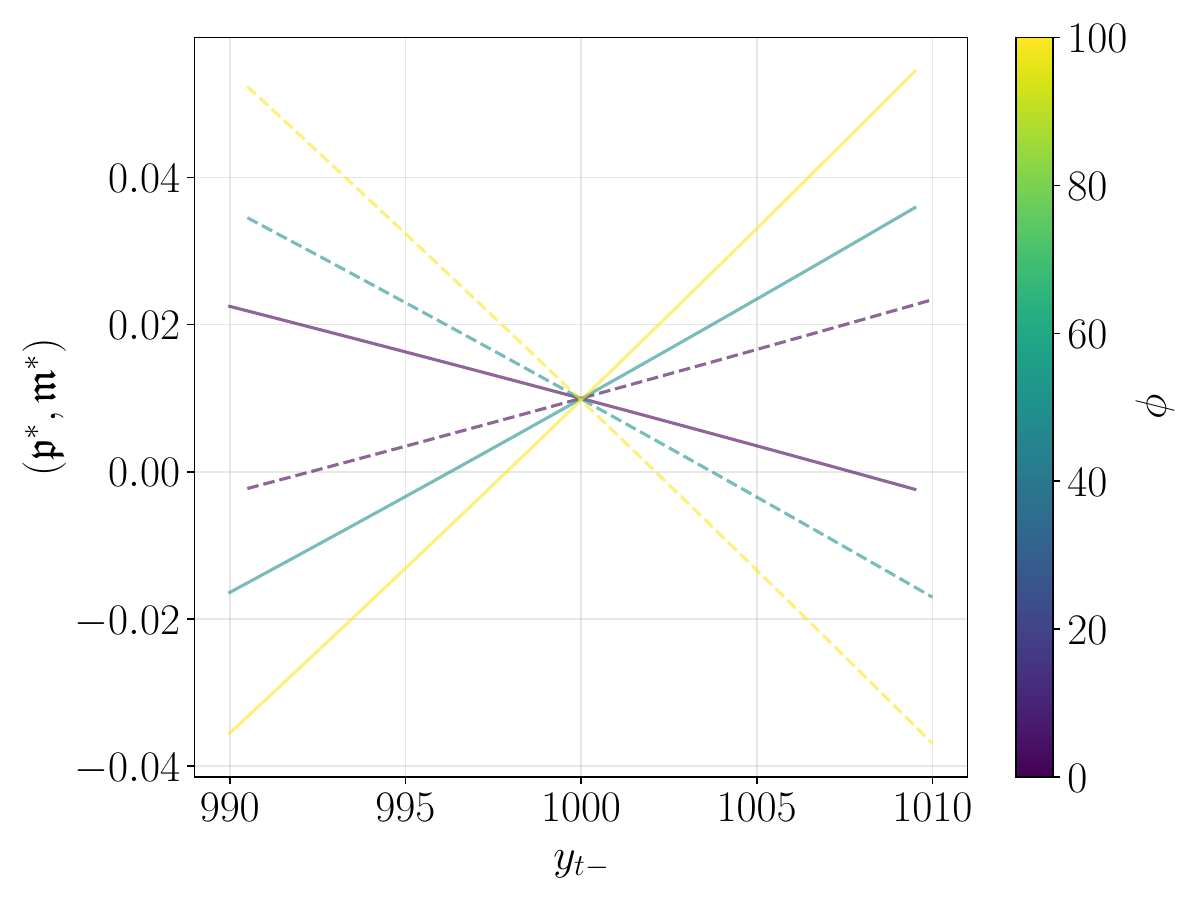}
        \caption{Optimal fees for selling $\feeLTsells^{*}(t,y_{t-})$ (solid line) and for buying $\feeLTbuys^{*}(t,y_{t-})$ (dashed line) at time $t=0.5$ as a function of the quantity of asset $Y$ in the pool and $\pencons$ (colorbar).}
        \label{fig: optimal fees second approx penalty}
    \end{subfigure}
    \hfill
    \begin{subfigure}[t]{0.49\textwidth}
        \centering
        \includegraphics[width=\textwidth]{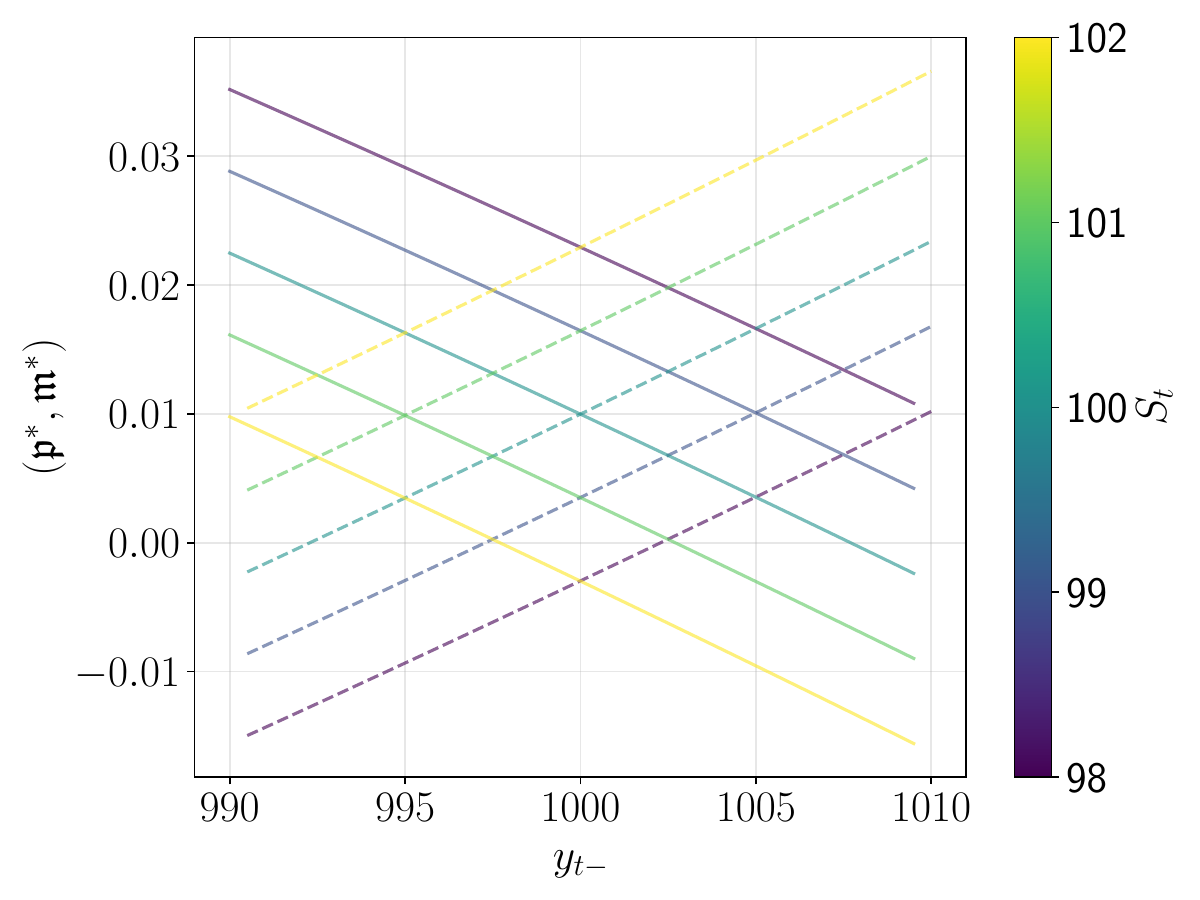}
        \caption{Optimal fees for selling $\feeLTsells^{*}(t,y_{t-})$ (solid line) and for buying $\feeLTbuys^{*}(t,y_{t-})$ (dashed line) at time $t=0.5$ as a function of the quantity of asset $Y$ in the pool and $\oracleprice$ (colorbar).}
        \label{fig: optimal fees second s}
    \end{subfigure}
\end{figure}

We now compare the value functions obtained from the two approximations. If $\sigma=0$ then the value function $v(t,y,\cash)$ (right) from Theorem \ref{th: value function and optimal fees} is the \emph{true} value function and we expect it to be close to the function $g(t,y,s)$ (left) in the case of $\sigma = 0$. The following figures shows the comparison between the plots for $g$ and $v$ depending from time and the quantity of asset $Y$. The interval $[0,1]$ has been discretized in $1000$ time steps.

\begin{figure}[H]
    \centering
    \begin{subfigure}[t]{0.31\textwidth}
        \centering
        \includegraphics[width=\textwidth]{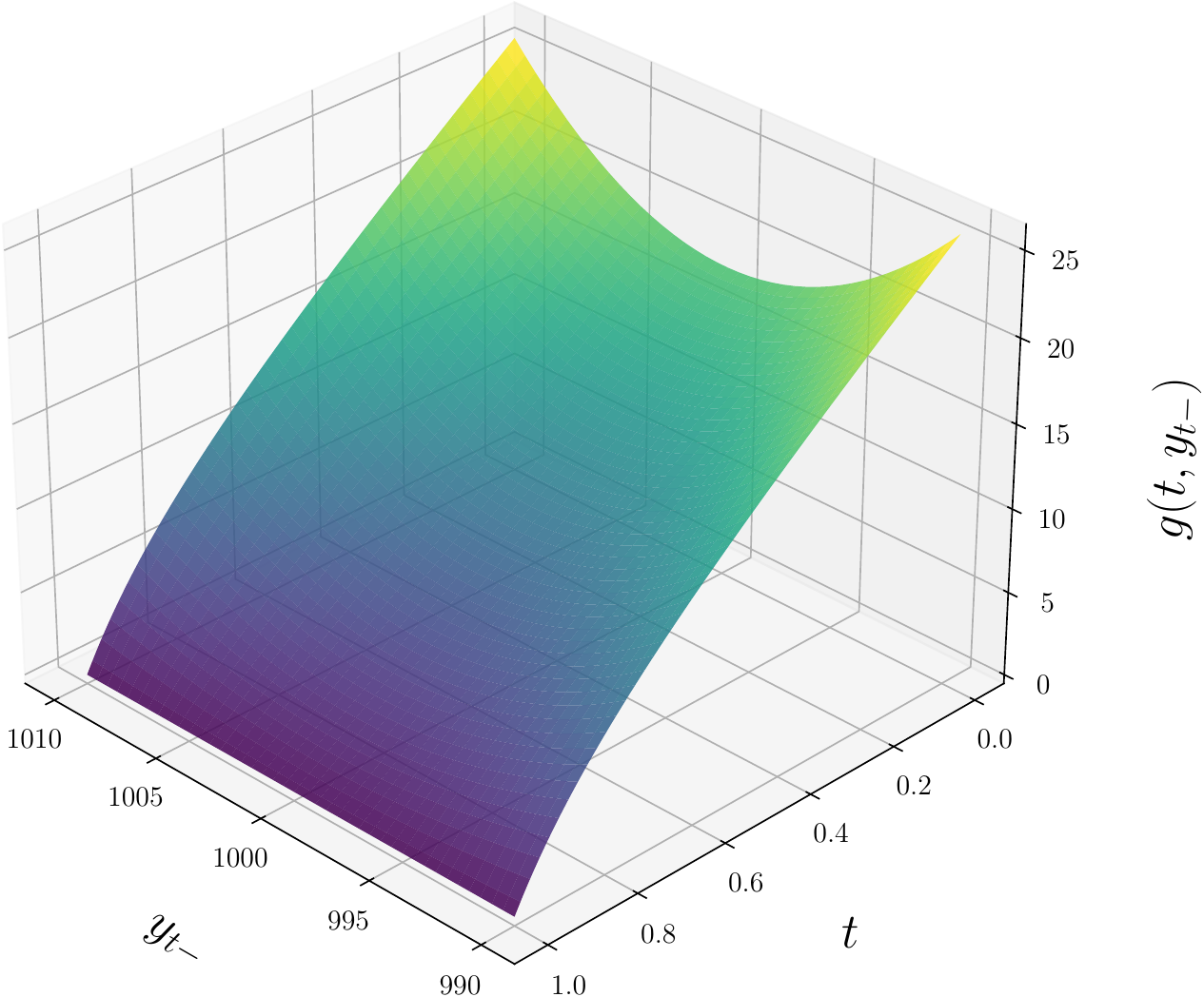}
        \caption{Value function first approximation.}
        \label{fig: value function first approx}
    \end{subfigure}
    \hspace*{0.01\textwidth}
    \begin{subfigure}[t]{0.31\textwidth}
        \centering
        \includegraphics[width=\textwidth]{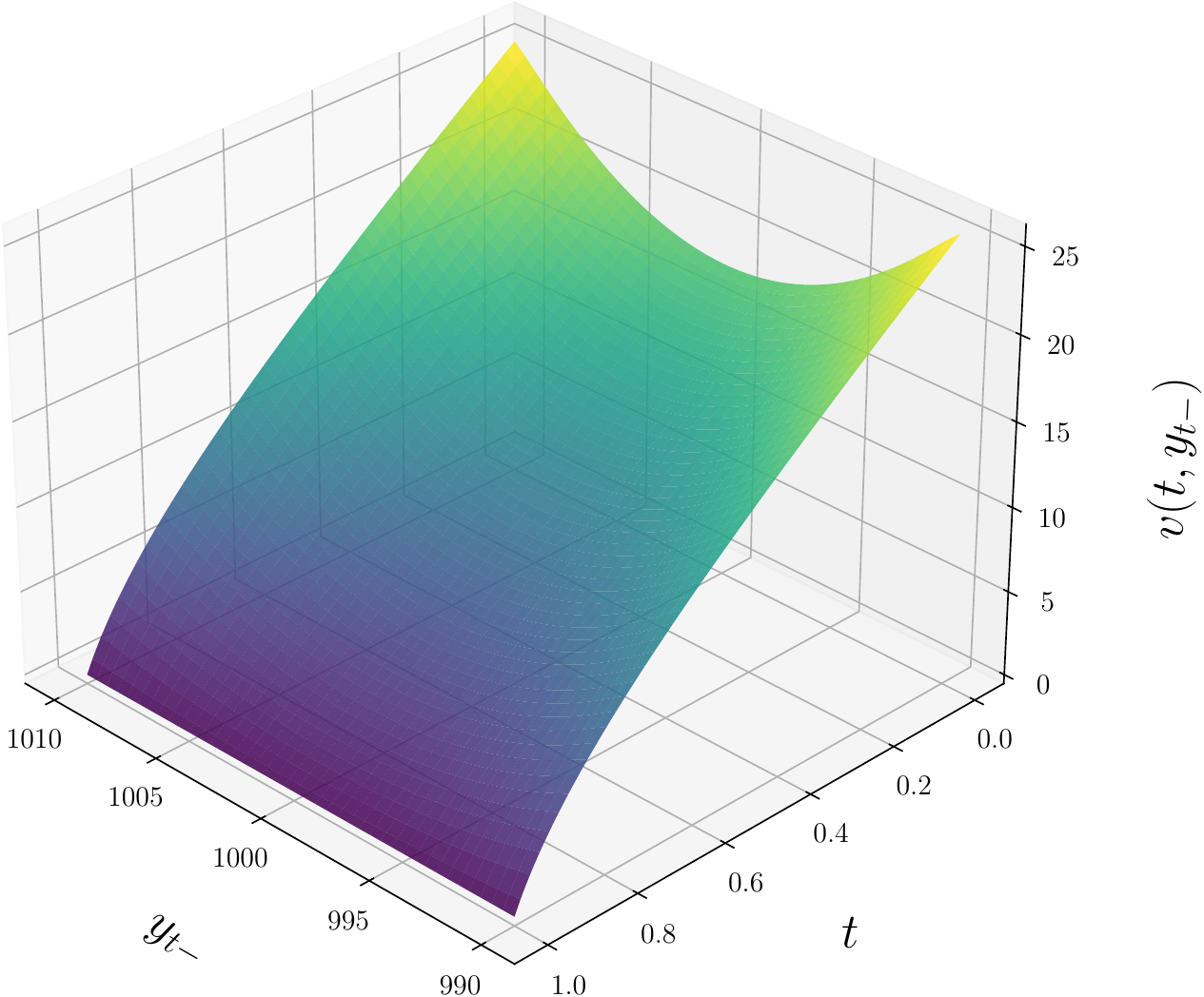}
        \caption{Second approximation when $\sigma = 0$.}
        \label{fig: value function second approx}
    \end{subfigure}
    \hspace*{0.01\textwidth}
    \begin{subfigure}[t]{0.31\textwidth}
        \centering
        \includegraphics[width=\textwidth]{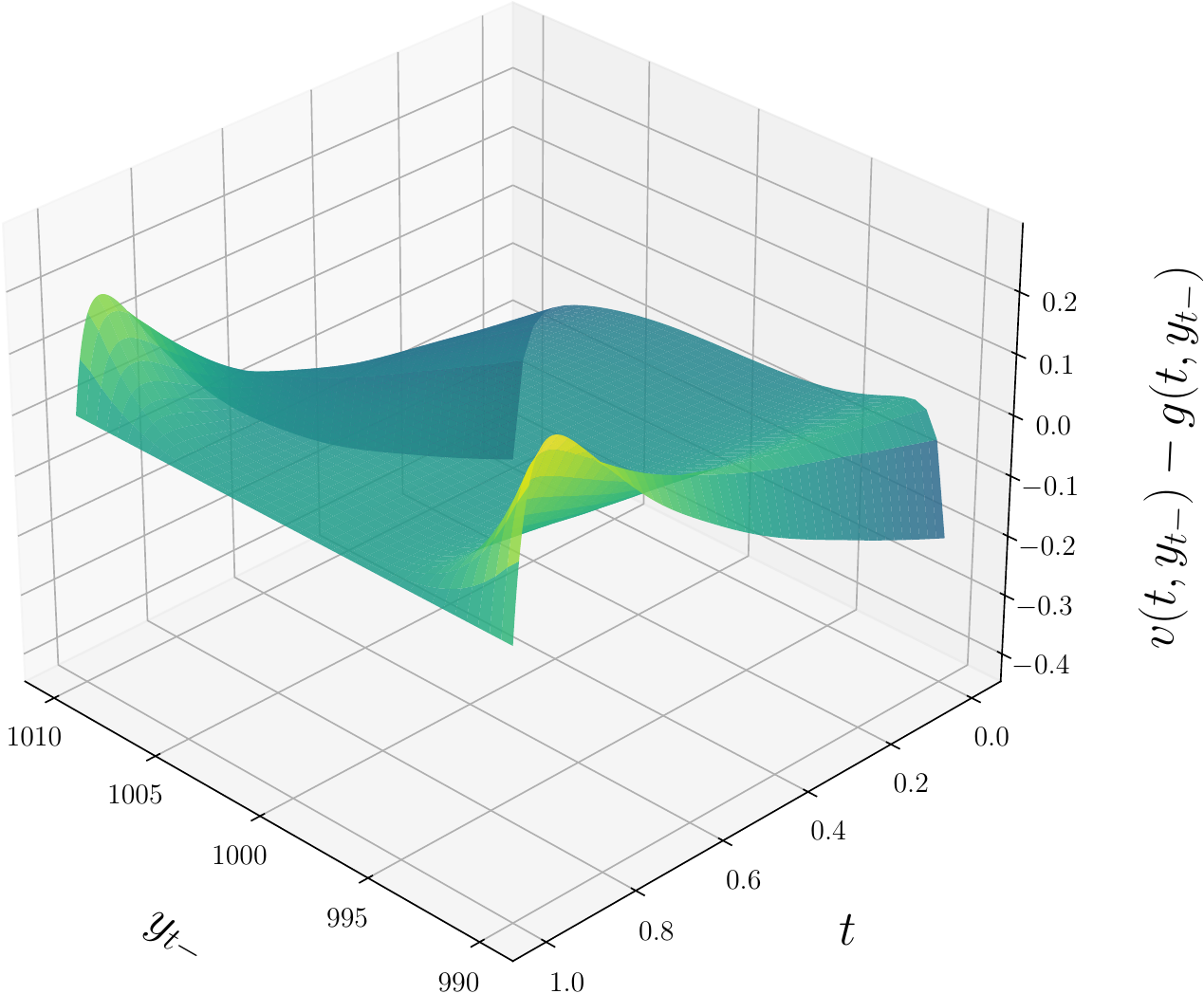}
        \caption{Difference between the two approximations.}
        \label{fig: difference}
    \end{subfigure}
\end{figure}

Finally we compare the optimal strategy from the second-order approximation in \eqref{thm: solution to the PDE system} (\emph{Optimal SA}) to the following strategies:
\begin{itemize}
    \item[$(i)$] the optimal strategy from the first approximation (\emph{Optimal FA}) implemented with a stochastic $S_t$ instead of $S_0$,
    \item[$(ii)$] the \emph{constant} strategy where the fees are constant for every time $t$ and every quantity $y$. The constant $c$ is chosen as the average of the optimal two fees at $t=0.5$ for $y = y_0$, i.e., \[ c = \frac{\feeLTsells^{*}(0.5,y_0) + \feeLTbuys^{*}(0.5,y_0)}{2}. \]
\end{itemize}

We use the following parameters: $\depth = 10^{8}$, $y_0 = 1000$, $\pencons = 0$, $\underline{y} = 990$, $\overline{y} = 1010$, $T=1$ and we run simulations for different $\intbuy$, $\intsell$ and $k$. We carry out 100,000 simulations and we discretise $[0,T]$ in $1,000$ timesteps.

In the following table the column \textit{fees} shows the revenue from collecting fees, the column \textit{sell} shows the number of sell orders, the column \textit{buy} shows the number of buy orders and the column \textit{QV} shows the quadratic variation of the instantaneous exchange rate $Z$ defined in Equation \ref{eq: marginal exchange}.

\begin{center}
\begin{tabular}[h]{@{}rcccccccc@{}}
\toprule
& \multicolumn{4}{c}{$\lambda^{+} = \lambda^{-} = 100$}
& \multicolumn{4}{c}{$\lambda^{+} = \lambda^{-} = 150$} \\
\cmidrule(lr){2-5} \cmidrule(lr){6-9}
& fees & sell & buy & QV
& fees & sell & buy & QV\\
\midrule

\multicolumn{9}{l}{$k=2$}\\
\midrule
Optimal SA   & 35.62 & 35.90 & 35.91 & 0.69 & 53.05 & 51.60 & 51.60 & 0.98 \\
Optimal FA    & 35.61 & 34.79 & 34.76 & 0.67 & 53.02 & 53.45 & 53.45 & 1.01 \\
Constant  & 35.18 & 35.18 & 35.19 & 0.68 & 52.29 & 52.28 & 52.29 & 0.99 \\

\midrule
\multicolumn{9}{l}{$k=1$}\\
\midrule
Optimal SA   & 71.49 & 35.92 & 35.93 & 0.69 & 106.42 & 53.46 & 53.47 & 1.00 \\
Optimal FA    & 71.51 & 35.47 & 35.44 & 0.68 & 106.38 & 53.46 & 53.51 & 1.01 \\
Constant & 71.23 & 35.60 & 35.61 & 0.69 & 105.90 & 52.91 & 52.91 & 1.00 \\
\bottomrule
\end{tabular}
\end{center}

As before, we find that either approximation yields a performance that is almost indistinguishable from the other, and both outperform constant fees substantially.
This supports the usage of linear dynamic fees when designing fee structures in AMMs.

\section{Dynamic Liquidity}\label{sec: Stochastic depth}

In this section, we extend our model to incorporate non-constant pool depth. This extension brings the framework closer to real-world AMMs, where liquidity is not constant over time. Indeed, liquidity providers may add or remove capital from the pool in response to market conditions, expected fee revenues, volatility, or changes in the attractiveness of alternative venues. Consequently, we treat the depth of the pool as a dynamic quantity rather than as an exogenous constant. In order to do this we assume that the depth of the pool takes values in a finite discrete grid
\begin{equation*} \label{eq: grid for p}
    \{p^{-M} : = \underline{p}, \dots, p^{0}, \dots, p^{M} : = \overline{p} \},
\end{equation*}
where $0 < \underline{p} < p^{0} < \overline{p} < \infty$. We denote by $\varrho^{+}$ and $\varrho^{-}$ the functions \[\varrho^{+}(p^{i}) := p^{i+1} - p^{i}, \quad \quad \varrho^{+}(p^{i}) := p^{i} - p^{i-1}. \]

Moreover, to explicitate the dependence from the depth we will denote with $f_{p^{i}} : \mathbb{R}_{+} \times \mathbb{R}_{+} \to \mathbb{R}_{+}$ the trading function such that $f_{p^{i}}(x,y) = (p^{i})^{2}$, where $x$ and $y$ denote the amounts in assets $X$ and $Y$ in the pool, respectively. Similarly, we explicitate the depth dependence on the level function denoting with $\varphi_{p^{i}}: \mathbb{R}_{+} \to \mathbb{R}_{+}$ the function such that $f(\varphi_{p^{i}}(y),y) = (p^{i})^{2}$. As a consequence we now have $2M+1$ (one for every different depth level) different grids for the risky asset $y$. We will denote the elements of the $i$-th grid for $y$ as

\begin{equation*}\label{eq: grid for quantity of y nonconstant liquidity}
         \{ y^{-N}_{p^{i}} := \underline{y_{p^{i}}}, \dots, y^{0}_{p^{i}}, \dots, y^{N}_{p^{i}} := \overline{y_{p^{i}}} \},
\end{equation*}
where $0 < \underline{y_{p^{i}}} < \overline{y_{p^{i}}}$.

Similar to the definitions above, we have:

\begin{enumerate}

    \item The \emph{marginal exchange rate} describes the price of an infinitesimal trade and is given by \begin{equation*}\label{eq: marginal exchange nonconstant liquidity}
        Z(i_{p},j_{y}) : = Z_{p^{i}} ( y^{j}_{p^{i}} ) : = - \varphi'_{p^{i}} (y^{j}_{p^{i}} ), \quad \quad i \in \{-M, \dots, M\}, \quad j \in \{-N, \dots ,N\}.
    \end{equation*}

    \item The \emph{exchange rate for buying} (taking out of the pool) $\deltabuy(i_{p},j_{y}) := \deltabuy_{p^{i}}(y^{j}_{p^{i}}):= y^j_{p^{i}}-y^{j-1}_{p^{i}}$ units of asset $Y$ is given by
    \begin{equation*}\label{eq: exchange rate for buying nonconstant liquidity}
        \exratebuy(i_{p},j_{y}) :  = \frac{\varphi_{p^{i}}(y^{j-1}) - \varphi_{p^{i}}(y^{j})}{\deltabuy(i_{p},j_{y})}, \quad \quad i \in \{-M, \dots, M\}, \quad j \in \{-N +1, \dots ,N\}.
    \end{equation*}
It takes values in the matrix $\mathcal{Z}^{-} \in  \mathbb{R}^{(2M +1) \times 2N}$, where
    \begin{equation*}\label{eq: matrix for buying exchange}
      (\mathcal{Z}^{-})_{i,j} : = \exratebuy(i_{p},j_{y}), \quad \quad i \in \{-M, \dots, M\}, \quad j \in \{-N +1, \dots ,N\}.
    \end{equation*}
    \item The \emph{exchange rate for selling} (depositing in the pool) $\deltasell(i_{p},i_{y}) := \deltasell_{p^{i}}(y^{i}):= y^{i+1}-y^{i}$ units of asset $Y$ is given by
    \begin{equation*}\label{eq: exchange rate for selling nonconstant liquidity}
        \exratesell(i_{p},j_{y}) : = \frac{\varphi_{p^{i}}(y^{j}) - \varphi_{p^{i}}(y^{j+1})}{\deltasell(i_{p},j_{y})}, \quad \quad i \in \{-M, \dots, M\}, \quad j \in \{-N, \dots ,N-1\}.
    \end{equation*}
    It takes values in the matrix $\mathcal{Z}^{+} \in  \mathbb{R}^{(2M +1) \times 2N}$
    \begin{equation*}\label{eq: matrix for selling exchange}
      (\mathcal{Z}^{+})_{i,j} : = Z_{+}(i_{p}, j_{y}), \quad \quad i \in \{-M, \dots, M\}, \quad j \in \{-N, \dots ,N-1\}.
    \end{equation*}
\end{enumerate}
Note that for the exchange rates for buying and selling satisfy the identity 
\begin{equation*}
\exratebuy(i_{p}, j_{y}) = \exratesell(i_{p}, j_{y} - 1 ), \quad \quad i \in \{-M, \dots, M\}, \quad j \in \{-N + 1, \dots ,N\}.
\end{equation*}

The order flow is modelled via the controlled point processes $\{ \ppbuy_{t} \}_{t \in [0,T]}$ and $ \{\ppsell_{t} \}_{t \in [0,T]}$, respectively, where the fee structure processes $\{\feeLTsells_{t}\}_{t \in [0, T]} $ and $\{\feeLTbuys_{t}\}_{t \in [0, T]}$ are assumed to be predictable.

The controls will operate in such a way that the difference $N_{t}^{+} - N_{t}^{-}$ takes values in $\{-N, \ldots, N\}$.
The quantity of asset $Y$ at time $t \in [0,T]$ is then given by
\begin{equation*}
Y_{t}^{\feeLTsells,\feeLTbuys}(P_{t}^{\feeLTsells,\feeLTbuys}) := y^{N_{t}^{+} - N_{t}^{-}}_{p_{t}}. 
\end{equation*}
For brevity we will omit the dependence of $Y_{t}^{\feeLTsells,\feeLTbuys}$ from $P_{t}^{\feeLTsells,\feeLTbuys}$ when it is clear.
Differently from before, the controlled intensities of $\{ N_{t}^{-} \}_{t \in [0,T]}$ and $ \{  N_{t}^{+} \}_{t \in [0,T]}$ are given by
\begin{align}\label{eq: intensities for LT ppp}
 \lambda_{t}^{-,\feeLTbuys} &: = \intbuy \exp{  \left( -\expdecay   (\exratebuy^{\feeLTbuys_{t}}(P_{t-}^{\feeLTsells,\feeLTbuys},Y_{t-}^{\feeLTsells,\feeLTbuys}) - (\oraclepricestochastic -\zeta)) \deltabuy(P_{t-}^{\feeLTsells,\feeLTbuys},Y_{t-}^{\feeLTsells,\feeLTbuys}) + \gamma P_{t}^{\feeLTsells,\feeLTbuys} \right) } \ind_{ \{ Y_{t-}^{\feeLTsells,\feeLTbuys} > \underline{y} \}}, \\  
   \lambda_{t}^{+,\feeLTsells} &: =  \intsell \exp{  \left( \expdecay( \exratesell^{\feeLTsells_t}(P_{t-}^{\feeLTsells,\feeLTbuys},Y_{t-}^{\feeLTsells,\feeLTbuys})  - (\oraclepricestochastic + \zeta)) \deltasell(P_{t-}^{\feeLTsells,\feeLTbuys},Y_{t-}^{\feeLTsells,\feeLTbuys}) + \gamma P_{t}^{\feeLTsells,\feeLTbuys} \right) } \ind_{ \{ Y_{t-}^{\feeLTsells,\feeLTbuys} < \overline{y} \}}.
\end{align}
Here, the two stochastic intensities are analogous to the previous model with the addition of the term $\gamma P_{t}^{\feeLTsells,\feeLTbuys}$. This term models the stylised fact that deeper pools attract more orders.

We now model how the depth behaves as a state variable. Change in liquidity is modelled via two Poisson processes $\{M^{+}_{t}\}_{t \in [0,T]}$ and $\{ M^{-}_{t} \}_{t \in [0,T]}$ for adding and subtracting liquidity, respectfully. We assume change in liquidity happens exogenously and hence these two processes have constant intensities $\eta^{+}$ and $\eta^{-}$.

The controls operates in such a way that the difference $M^{+}_{t} - M^{-}_{t}$ takes value in $\{-M, \dots, M\}$. The depth of the pool $P$ at time $t \in [0,T]$ is then given by \[ P_{t} : = p^{M^{+}_{t} - M^{-}_{t}} .\]
The AMM seeks to solve the control problem \[ v(t,\cash,y,p,s) : = \sup_{(\feeLTbuys,\feeLTsells) \in \mathcal{A}_{t}} v^{(\feeLTbuys,\feeLTsells)}(t,\cash,y,p,s),\]
where $\mathcal{A}_{t}$ denotes the set of all $\mathbb{F}$-predictable and bounded fee structure processes $(\feeLTbuys_{u},\feeLTsells_{u})_{\{t \leq u \leq T\}}$ and the conditional performance criterion is given by \[ v^{(\feeLTbuys,\feeLTsells)}(t,\cash,y,p,s) : = \mathbb{E}_{(t,\cash,y,p,s)} \left[ \Cash_{T}^{(t,\cash,y,p,s,\feeLTbuys,\feeLTsells)} - \int_{t}^{T}  \Pi(Y^{(t,\cash,y,p,s,\feeLTbuys,\feeLTsells)}_{u},S_{u}^{(t,\cash,y,p,s)}) \dd u \right]. \]
Here, $\{ \Cash_{u}^{(t,\cash,y,p,s,\feeLTbuys,\feeLTsells)} \}_{u \in [t,T]}$, $\{ Y_{u}^{(t,\cash,y,p,s,\feeLTbuys,\feeLTsells)} \}_{u \in [t,T]}$, $\{ P_{u}^{(t,\cash,y,p,s,\feeLTbuys,\feeLTsells)} \}_{u \in [t,T]}$ and $\{ S_{u}^{(t,\cash,y,p,s)}\}_{u \in [t,T]} $ denote the (controlled) processes $\Cash$, $Y$, $P$ and $S$ restarted at time $t$ with initial value $\cash$, $y$, $p$ and $s$, respectively.

From the dynamic programming principle, we determine that the Hamilton-Jacobi-Bellman (HJB) equation satisfied by the value function is:

\begin{equation}
    \begin{aligned}
     0=\;&\frac{\partial}{\partial t} v(t,\cash,y,p,s)
+\frac{\sigma^{2}}{2}\,\frac{\partial^{2}}{\partial s^{2}} v(t,\cash,y,p,s)
- P(y,p,s) \\[2pt]
&\;+\;\eta^{+}\Big[v\big(t,\cash,y,\,p+\varrho^{+}(p,y),\,s\big)-v(t,\cash,y,p,s)\Big]\;\ind_{\{p<\overline p\}} \\
&\;+\; \eta^{-}\Big[v\big(t,\cash,y,\,p-\varrho^{-}(p,y),\,s\big)-v(t,\cash,y,p,s)\Big]\;\ind_{\{p>\underline p\}}\\[4pt]
&\;+\;\Bigg(\intsell e^{\expdecay(\exratesell(p,y)-\oracleprice)\,\deltasell(p,y) + \gamma p}\;
\sup_{\feeLTsells\in\mathbb{R}} e^{-\expdecay\,\feeLTsells\,\exratesell(p,y)\,\deltasell(p,y)}\\[-2pt]
&\qquad\qquad\qquad\qquad\Big[v\big(t,\cash+\feeLTsells\,\exratesell(p,y)\,\deltasell(p,y),\,y+\deltasell(p,y),\,p,\,s\big)
- v(t,\cash,y,p,s)\Big]\Bigg)\;\ind_{\{y<\overline y\}}\\[4pt]
&\;+\;\Bigg(\intbuy e^{-\expdecay(\exratebuy(p,y)-\oracleprice)\,\deltabuy(p,y) + \gamma p }\;
\sup_{\feeLTbuys\in\mathbb{R}} e^{-\expdecay\,\feeLTbuys\,\exratebuy(p,y)\,\deltabuy(p,y)}\\[-2pt]
&\qquad\qquad\qquad\qquad\Big[v\big(t,\cash+\feeLTbuys\,\exratebuy(p,y)\,\deltabuy(p,y),\,y-\deltabuy(p,y),\,p,\,s\big)
- v(t,\cash,y,p,s)\Big]\Bigg)\;\ind_{\{y>\underline y\}}.   
    \end{aligned}
\end{equation}

Now the analysis carries on verbatim from the case with constant depth and we get

\begin{align}\label{eq: eq for g nonconstant liquidity}
    & \frac{\partial}{\partial t}  g(t,y,p) + \frac{\sigma^{2}}{2} \frac{\partial^{2}}{\partial s^{2}} g(t,y,p) - \Pi(y,p,s)  \nonumber \\
    &\;+\;\eta^{+}\Big[ g\big(t,y,\,p+\varrho^{+}(p) \big)-g(t,y,p)\Big]\;\ind_{\{p<\overline p\}}
    + \eta^{-}\Big[g\big(t,y,\,p-\varrho^{-}(p)\big)-g(t,y,p)\Big]\;\ind_{\{p>\underline p\}}\\[4pt]
    & \quad + \left( \frac{\intsell e^{- \expdecay \oracleprice \deltasell(p,y) - 1 + \gamma p}}{\expdecay} e^{\expdecay \exratesell (p,y)  \deltasell(p,y)} e^{ \expdecay ( g(t,y + \deltasell(p,y),s) - g(t,y,s) )} \right) \ind_{ \{ y < \overline{y} \}}  \\
    &\quad + \left( \frac{\intbuy e^{ \expdecay \oracleprice \deltabuy(p,y) - 1 + \gamma p}}{\expdecay} e^{-\expdecay \exratebuy (p,y)  \deltabuy(p,y)} e^{ \expdecay ( g(t,y - \deltabuy(p,y),s) - g(t,y,s))} \right) \ind_{ \{ y > \underline{y} \}} = 0. \nonumber
\end{align}

Since fees are updated at high frequency, the pool depth is unlikely to experience large shocks between two consecutive fee adjustments. 
We therefore treat liquidity variations as having only a negligible first-order effect on the continuation value and approximate\footnote{To assess the accuracy of the approximation, we also compute the value function numerically using a backward Euler scheme. Our numerical results show that the approximation remains very accurate provided that the depth does not experience large jumps at each liquidity update. All the code used to generate the numerical results is available in the public GitHub repository of the project.} the value function as locally non-reactive to changes in depth: 

\[ g\big(t,y,\,p+\varrho^{+}(p) \big)-g(t,y,p) \approx 0, \quad \quad g\big(t,y,\,p-\varrho^{-}(p)\big)-g(t,y,p) \approx 0. \]

The final result is summarised in the following theorem.

\begin{theorem} \label{th: value function and optimal fees nonconstant liquidity}
    Fix a depth level $p^{h}$. Define the matrix $ \mathbf{A}(p^{h}) : =  ( \mathbf{A}_{i,j} (p^{h}) )_{0 \leq i \leq j \leq 2N} $ by
    \[
    \mathbf{A}_{i,j} (p^{h}) : =
    \begin{cases}
        - \expdecay \Pi (y^{j - N},p^{h},s)  & \text{ if } i = j, \\
        \intsell e^{- \expdecay \oracleprice \deltasell(p^{h},y^{j - N}) - 1 + \gamma p^{h} } e^{\expdecay \exratesell (p^{h},y^{j - N})  \deltasell(p^{h},y^{j - N})} & \text{ if } i = j - 1, \\
        \intbuy e^{ \expdecay \oracleprice \deltabuy(p^{h},y^{j - N}) - 1 + \gamma p^{h}}e^{-\expdecay \exratebuy (p^{h},y^{j - N})  \deltabuy(p^{h},y^{j - N})} & \text{ if } i = j + 1 \\
        0 & \text{ otherwise,}
    \end{cases}
    \]
    and denote with $\mathbf{1}$ the unit vector of $\mathbb{R}^{2N}$. Define the function $w : [0,T] \times \{-M, \dots, M \} \times \{ -N, \dots, N \} \to \mathbb{R}$ by \[ w(t,h,i) : = (\exp( \mathbf{A}(p^{h})(T - t)) \mathbf{1})_{i}, \]
    and the function $v : [0,T] \times \{-M, \dots, M \} \times \{ -N, \dots, N \} \times \mathbb{R}_{+} \to \mathbb{R}$ as
    \[ v(t,h,i,\cash) = \cash + \frac{1}{k}\log(w(t,h,i)).\]
    Then $v$ solves the Hamilton-Jacobi-Bellman equation
\begin{align*}
    & \frac{\partial}{\partial t} v(t,h,i,\cash) - \Pi(h,i,s) + \nonumber \\
    & \left( \intsell e^{- \expdecay \oracleprice \deltasell(h,i)} \sup_{ \feeLTsells \in \mathbb{R}} e^{\expdecay \exratesell^{\feeLTsells}(h,i) \deltasell(h,i)} \left[v(t,i+1, \cash + \feeLTsells \exratesell(h,i) \deltasell(h,i)) - v(t,h,i,\cash)\right] \right) \ind_{ \{ i < N \}} \\
    & \left(  \intbuy e^{\expdecay \oracleprice \deltabuy(h,i)} \sup_{ \feeLTbuys \in \mathbb{R}} e^{-\expdecay \exratebuy^{\feeLTbuys}(h,i) \deltabuy(h,i)} \left[v(t,i-1,\cash +  \feeLTbuys \exratebuy(h,i) \deltabuy(h,i)) - v(t,h,i,\cash) \right] \right) \ind_{ \{ i > -N \}} = 0, \nonumber
\end{align*}
with boundary condition $v(T,h,i,\cash) = \cash$ for every $(h,i) \in \{-M, \dots, M \} \times \{ -N, \dots, N \}$.  Moreover, the corresponding optimizers are independent of $\cash$ and satisfy

\begin{align}
    \feeLTsells^{*}(t,h,i) & : =  \frac{1}{\expdecay \exratesell(h,i) \deltasell(h,i)} \left( 1 +  \log \left( \frac{w(t,h,i)}{w(t,h,i+1)} \right) \right), \\
    \feeLTbuys^{*}(t,h,i) & : = \frac{1}{\expdecay \exratebuy(h,i) \deltabuy(h,i)} \left( 1 +  \log \left( \frac{w(t,h,i)}{w(t,h,i-1)} \right) \right),
\end{align}
for $(h,i) \in \{-M, \dots, M \} \times \{ -N, \dots, N \}$.
\end{theorem}

\subsection{Numerical results for stochastic liquidity}\label{sec:numerical_results_nonconstant}

In this section we present numerical simulations of the optimal fee structure under the non-constant liquidity framework. We perform our analysis on a constant product market maker (CPM) where the trading function is $f(x,y) = xy$ and the level function with depth $p^{2}$ is $\varphi(y)_{p} = p^2/y$. 
\newline
In the numerical experiments of the appendix, we consider a pool with trading horizon \(T=1\), baseline intensities \(\lambda^-=\lambda^+=50\), exponential decay parameter \(k=2\), centralised reference price \(S_0=100\), and liquidity penalty constant \(\gamma=10^{-20}\). The reference pool depth is set to \(p_0^2=10^8\), and the initial risky-asset inventory is chosen so that the pool price is aligned with the centralised reference price, namely \(y_0=1000\). The depth variable is discretized into seven levels,
\[
p^2 \in \left\{1.25\times 10^7,\;2.50\times 10^7,\;5.00\times 10^7,\;1.00\times 10^8,\;2.00\times 10^8,\;4.00\times 10^8,\;8.00\times 10^8\right\},
\]
with the central value \(p_0^2=10^8\) serving as the reference depth. Thus, the grid contains both shallower and deeper pools around the benchmark level, with adjacent grid points differing by a factor of two.

\begin{figure}[H]
    \centering
    \includegraphics[width=0.9\textwidth]{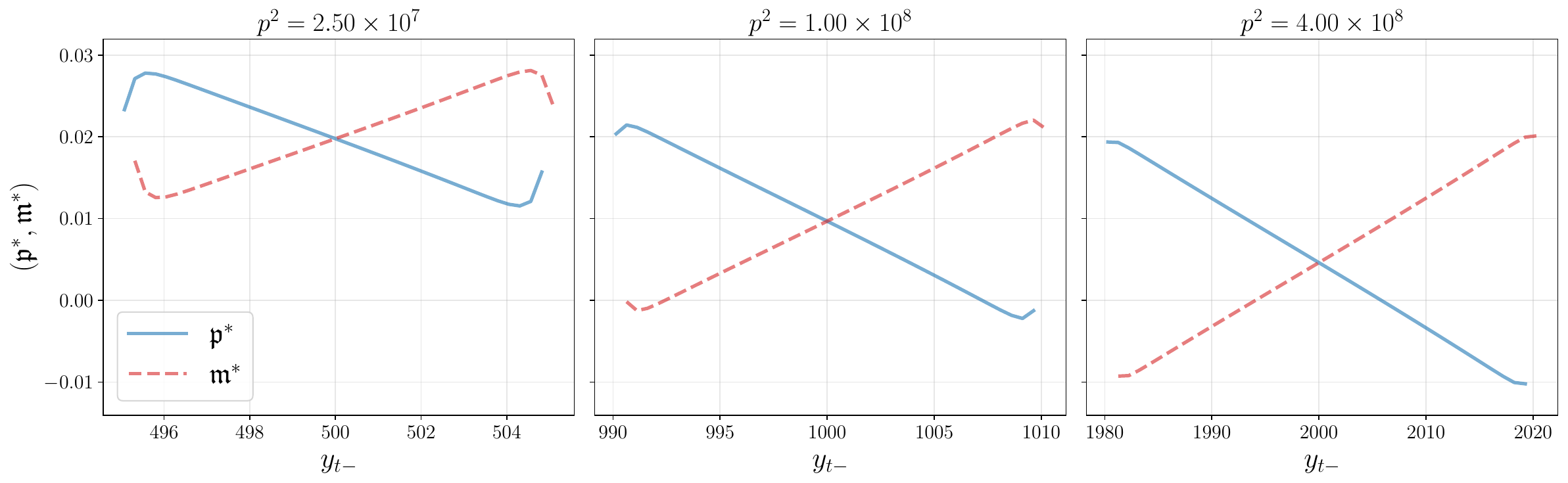}
    \caption{Optimal selling fees $\mathfrak{p}^*(t,y)$ (solid line) and buying fees $\mathfrak{m}^*(t,y)$ (dashed line) as a function of the risky asset $y$ at $t=0.5$ across three liquidity depth levels.}\label{fig:optimal_fees_depths}
\end{figure}

For each fixed depth level \(p^2\), we then construct a grid for the risky-asset inventory \(y\) so that the corresponding marginal exchange rate \(Z(y)\) changes by \(0.1\) between neighboring grid points. More precisely, for a given reference inventory \(y_0\), the grid is defined by
\[
y^i := \sqrt{\frac{p^2}{Z(y_0)-0.1\,i}},
\qquad i\in\{-20,\dots,20\}.
\]
In this way, the inventory grid is not chosen uniformly in \(y\), but rather implicitly through an equidistant grid in price space. This ensures that each step in the \(y\)-grid corresponds to the same change in the pool exchange rate, which is convenient for the numerical implementation and facilitates comparisons across different depth levels. Figure~\ref{fig:optimal_fees_depths} displays the optimal fee structure \(\mathfrak{p}^*(t,y)\) and \(\mathfrak{m}^*(t,y)\) at three representative depth levels: a shallow pool \((h = 1.25 \times 10^7)\), the reference pool \((h = 1.00 \times 10^8)\), and a deep pool \((h = 8.00 \times 10^8)\). As a function of the inventory variable \(y\), the optimal fees exhibit the same qualitative behavior as in the constant-liquidity case, with two distinct regimes depending on whether the pool inventory lies below or above the reference level. The main additional feature is that the magnitude of the optimal fees decreases substantially as the liquidity depth increases.

This monotonicity can be explained directly from the structure of the model. Since the inventory grid is constructed through an equidistant grid in exchange-rate space, the relevant price levels remain comparable across depth levels, whereas the inventory increments increase with the depth parameter. By the characterization of the optimal fees, the latter are inversely proportional to the product of the local exchange rate and the corresponding inventory increment. Therefore, as the depth increases, the denominator in the fee formulas becomes larger, which mechanically leads to smaller optimal fees. In the present calibration, the depth-sensitivity parameter is small, so the dependence of the continuation value on depth is dominated by this geometric scaling effect. As a result, deeper pools optimally charge lower fees. The economic intuition is that, when pool depth is higher, a larger trading volume is required to move the AMM quote and realign it with the external price. 
As a result, the venue charges lower percentage fees while still collecting substantial revenues in absolute terms. 
In this sense, deeper pools can afford to be more competitive on fees because they benefit from larger trade sizes.

A second effect visible in Figure~\ref{fig:optimal_fees_depths} is that, when the fees are plotted as functions of the inventory variable \(y\), the curves appear more compressed for smaller depth levels and more spread out for larger depth levels. This is a direct consequence of the construction of the inventory grid. Indeed, for each depth level, the grid is defined by
\[
y^{i}_{p^{j}}:=\sqrt{\frac{(p^{j})^{2}}{Z(y_{0})-0.1\,i}},
\qquad i\in\{-N^{j},\dots,N^{j}\}.
\]
so that, for fixed \(i\), the inventory values scale proportionally with the depth parameter \(p\). Hence, the same equidistant discretization in exchange-rate space corresponds to a narrower interval in \(y\) for shallow pools and to a wider interval in \(y\) for deeper pools. Therefore, when represented as functions of \(y\), the fee profiles are horizontally compressed at low depth and horizontally stretched at high depth.

\begin{figure}[H]
    \centering
    \includegraphics[width=0.6\textwidth]{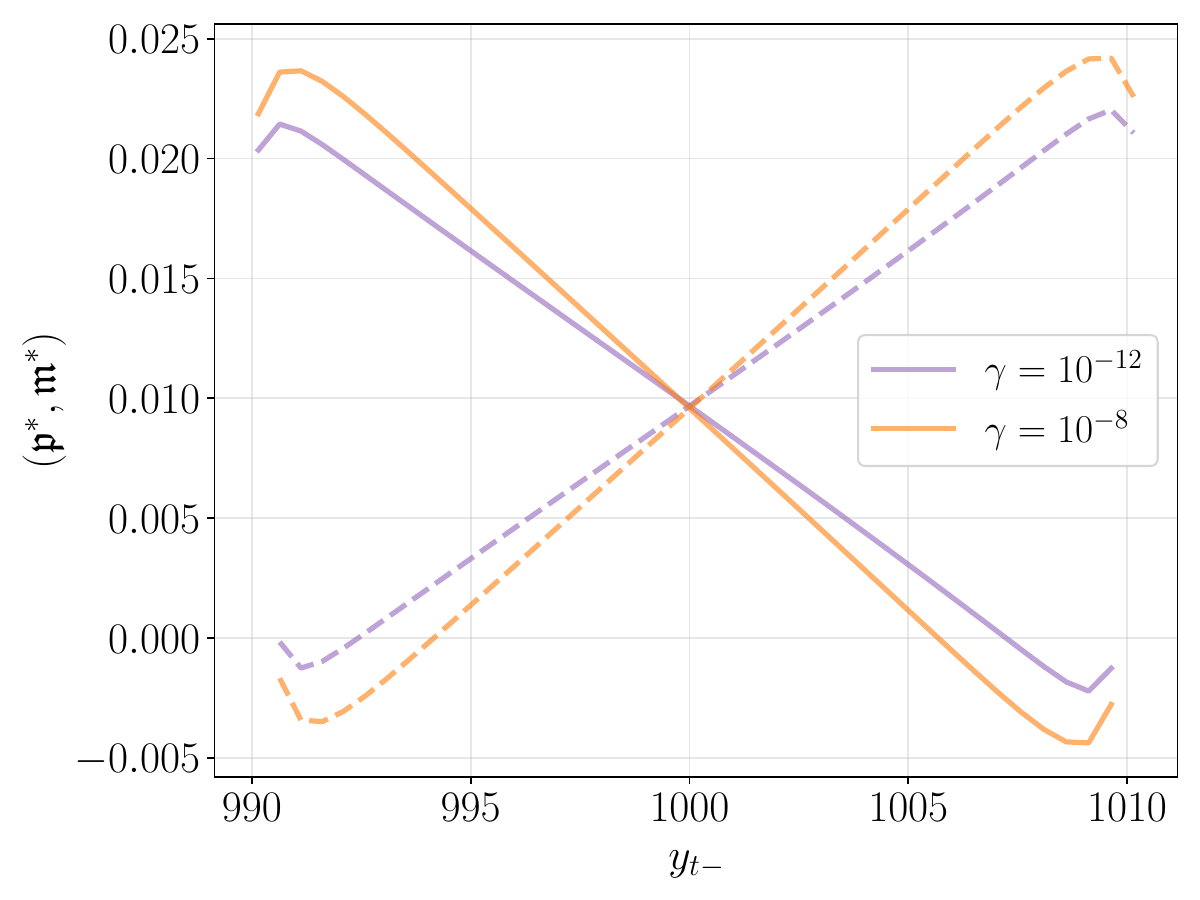}
    \caption{Optimal fees at the reference depth ($h = 10^8$) for two values of the liquidity penalty parameter $\gamma$. Selling fees $\mathfrak{p}^*(t,y)$ are shown as solid lines; buying fees $\mathfrak{m}^*(t,y)$ are shown as dashed lines.}\label{fig:optimal_fees_gamma}
\end{figure}

To understand how exogenous factors affect fee dynamics, Figure~\ref{fig:optimal_fees_gamma} compares optimal fees under two values of the liquidity penalty parameter $\gamma$: $\gamma = 10^{-12}$ and $\gamma = 10^{-8}$. The parameter $\gamma$ controls the sensitivity of order arrival intensities to the pool depth.

The results show that larger values of \(\gamma\) lead to more pronounced fee curves. This can be interpreted from the fact that \(\gamma\) governs how strongly order arrival intensities respond to the pool depth. When \(\gamma\) is large, deeper liquidity provides a stronger support to order flow, so the AMM can adjust fees more aggressively without reducing trading activity too severely. In this case, the trade-off between fee extraction and order-flow preservation is weaker, which results in steeper optimal fee schedules. Conversely, when \(\gamma\) is small, order flow is less sensitive to depth, so fees have a more direct impact on trading intensity. The AMM must then moderate its fee schedule in order to avoid discouraging too much order flow. Hence the optimal fees are flatter when \(\gamma\) is low.

We now compare the performance of three fee strategies across four parameter combinations, evaluated via Monte Carlo simulation with \(100{,}000\) paths per strategy. We compare our optimal strategy with two alternative strategies:
\begin{itemize}
    \item[$(i)$] the linear strategy, where at each time the fees are given by the linearized controls \(\feeLTsells^{*}(t,y)\) and \(\feeLTbuys^{*}(t,y)\) in a neighbourhood of \(y_0\), and
    \item[$(ii)$] the \emph{constant} strategy, where the fees are constant for every time \(t\), every quantity \(y\), and every depth level \(p\). The constant \(c\) is chosen as the average of the two optimal fees at time \(t=0.5\), evaluated at the reference depth level \(p_0\) and at the corresponding central inventory point \(y^{0}_{p_0}\), that is,
    \[
    c=\frac{\feeLTsells^{*}(0.5,y^{0}_{p_0})+\feeLTbuys^{*}(0.5,y^{0}_{p_0})}{2}.
    \]
\end{itemize}

In the following table, the column \textit{fees} shows the revenue from collecting fees, the column \textit{sell} shows the number of sell orders, the column \textit{buy} shows the number of buy orders.
 
\begin{center}

\begin{tabular}{@{}lcccccc@{}}

\toprule

& \multicolumn{3}{c}{$\lambda^{+} = \lambda^{-} = 100$}

& \multicolumn{3}{c}{$\lambda^{+} = \lambda^{-} = 150$} \\

\cmidrule(lr){2-4} \cmidrule(lr){5-7}

& fees & sell & buy

& fees & sell & buy \\

\midrule

\multicolumn{7}{l}{$k=2$}\\

\midrule

Optimal  & 35.6 & 35.9 & 35.9 & 53.1 & 53.5 & 53.5 \\

Linear   & 35.6 & 35.9 & 36.0 & 53.1 & 53.5 & 53.5 \\

Constant & 28.7 & 36.8 & 36.8 & 42.0 & 54.7 & 54.7 \\

\midrule

\multicolumn{7}{l}{$k=1$}\\

\midrule

Optimal  & 71.6 & 36.0 & 36.0 & 106.5 & 53.5 & 53.5 \\

Linear   & 71.6 & 36.0 & 36.0 & 106.5 & 53.5 & 53.5 \\

Constant & 57.7 & 36.5 & 36.5 & 85.8 & 54.1 & 54.1 \\

\bottomrule

\end{tabular}

\end{center}

As in the constant-liquidity case, the Monte Carlo results show that the optimal dynamic fees and their linear approximation are essentially indistinguishable in terms of average revenue. Across all parameter configurations, the difference between the two strategies is numerically negligible, which confirms that the linear rule remains an accurate and practically useful approximation of the fully optimal policy.

By contrast, the constant-fee strategy performs substantially worse in the present setting. The reason is that, once both the risky-asset inventory and the pool depth evolve over time, fees must adapt to changing market conditions. A constant fee cannot react either to inventory imbalances or to variations in liquidity, and therefore fails to adjust efficiently when the state of the pool moves away from its reference region. This effect is much stronger than in the constant-liquidity case.

The magnitude of the underperformance depends on the parameters. Relative to the dynamic strategies, the constant-fee policy generates approximately \(19.3\%\) less revenue when \(k=1\) and \(\lambda=100\), and about \(19.4\%\) less when \(k=1\) and \(\lambda=150\). For \(k=2\), the revenue loss is approximately \(19.5\%\) for \(\lambda=100\) and \(19.8\%\) for \(\lambda=150\). Hence, the gap remains consistently close to \(20\%\) across specifications and increases slightly as \(\lambda\) increases.

At the same time, the constant fee remains informative as an average benchmark. Since it is constructed from the optimal fees at the reference inventory level, it provides a reasonable description of fees in the central region of the state space. However, it is no longer optimal when the market changes state, either because the risky-asset inventory moves far from its reference value or because the pool depth approaches the boundaries of the depth grid. In those situations, state-dependent fee adjustment becomes crucial.

\begin{table}[H]
\centering
\caption*{}
\ra{1.2}
\begin{tabular}{@{}lccc ccc@{}}
\toprule
& \multicolumn{3}{c}{$\lambda^{+} = \lambda^{-} = 100$}
& \multicolumn{3}{c}{$\lambda^{+} = \lambda^{-} = 150$} \\
\cmidrule(lr){2-4} \cmidrule(lr){5-7}
& $\mathfrak{p}^*(0.5,y_0)$ & $\mathfrak{m}^*(0.5,y_0)$ & $c^*$
& $\mathfrak{p}^*(0.5,y_0)$ & $\mathfrak{m}^*(0.5,y_0)$ & $c^*$ \\
\midrule

\multicolumn{7}{l}{$k=2$}\\
\midrule
& 0.008816 & 0.010388 & 0.009602
& 0.008724 & 0.010419 & 0.009571 \\

\midrule

\multicolumn{7}{l}{$k=1$}\\
\midrule
& 0.019021 & 0.020311 & 0.019666
& 0.018889 & 0.020356 & 0.019623 \\
\bottomrule
\end{tabular}
\end{table}

The average constant fees used in the simulations are reported in the table above. As in the constant-liquidity case, these values depend strongly on \(k\): when \(k\) decreases from \(2\) to \(1\), the optimal constant fee roughly doubles. This reflects the fact that, when order arrivals are less sensitive to price deviations, the venue can sustain higher fee levels. Nevertheless, even with this calibration, the constant strategy remains clearly dominated by the dynamic ones in the variable-liquidity model.

\subsection{Equilibrium liquidity}
\begin{figure}[H]
    \centering
    \includegraphics[width=0.6\textwidth]{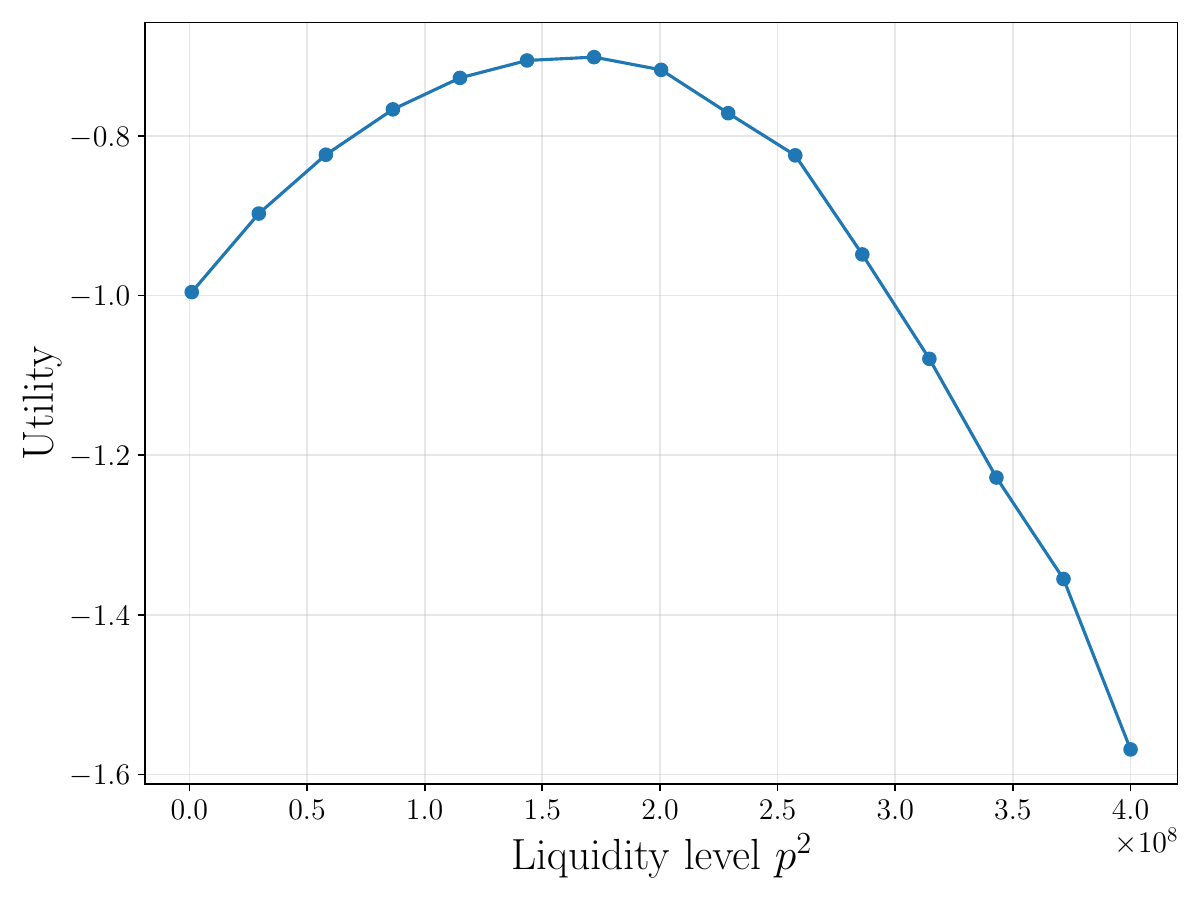}
    \caption{Exponential utility of impermanent loss and fees.}\label{fig:equilibrium liquidity}
\end{figure}

The optimal fees throughout this paper are computed for a given amount of initial liquidity $p^2$. If follows that the optimal feedback fee schedules we derived depend on $p^2$, more precisely, for any initial liquidity $p^2$, the optimal feedback maps are of the form $\mathfrak{p}^*(t,y;p^2)$ and $\mathfrak{m}^*(t,y;p^2)$ which we write as $\mathfrak{p}^*(p^2)$, $\mathfrak{m}^*(p^2)$ for notational simplicity. Thus,  we think of the equilibrium liquidity\footnote{This is similar to the core idea in  \cite{aqsha2025equilibriumrewardliquidityproviders}. } for such a fee schedule as the solution to the optimisation problem of the form
\begin{equation}
    \sup_{p^2\in (0,\bar{p}^2]} \mathbb{E}^{\mathfrak{p}^*(p^2),\mathfrak{m}^*(p^2)}\bigg[- e^{-\gamma( \mathfrak{f}\,\mathfrak{C}_T- \mathrm{IL}_T)}\bigg],
\end{equation}
where $\mathrm{IL}_T = \varphi(Y^{\mathfrak{p},\mathfrak{m}}_T) -\varphi(Y^{\mathfrak{p},\mathfrak{m}}_0) + S_t\,(Y^{\mathfrak{p},\mathfrak{m}}_T - Y^{\mathfrak{p},\mathfrak{m}}_0)$ is the impermanent loss, $\mathfrak{f}$ is the percentage of the fees collected that the venue transfer to their LPs, and $\mathbb{E}^{\mathfrak{p}^*(p^2),\mathfrak{m}^*(p^2)}$ stands for the expectation when the fee schedule is given by the optimal response to the liquidity level $p^2$.
 Figure \ref{fig:equilibrium liquidity} shows the result of such an optimisation problem for the baseline values of model parameters, $\gamma = 10^{-20}$, and $\mathfrak{f}=1$ for simplicity. 

We see that for this region of the parameter space there is a clear equilibrium liquidity  between $1.5\times 10^8$ and $2\times 10^8$. This could be thought as the equilibrium response to the venue's optimal fee schedules.

\section{Conclusion}

In this paper, we study how an AMM maximizes its revenue by setting optimal dynamic trading fees. We find that the optimal fee schedule serves two distinct purposes. On the one hand, by monitoring external prices, the AMM raises fees strategically to penalize arbitrageurs. On the other hand, it lowers fees to increase price volatility, which attracts noise traders and stimulates trading volume. This dual approach allows the AMM to balance the trade-off between deterring arbitrageurs and encouraging liquidity taking through noise order flow.
Future research could allow liquidity providers to act strategically, making liquidity provision endogenous rather than fixed. The model could also incorporate gas costs, delayed reference prices, and other market frictions. Finally, extending the framework to concentrated-liquidity AMMs would improve its applicability to modern decentralized exchanges.

\bibliography{thebib}
\bibliographystyle{plainnat}

\end{document}

%% file: macros.tex
\newcommand{\feeLTsells}{\mathfrak{p}}
\newcommand{\feeLTbuys}{\mathfrak{m}}
\newcommand{\exratesell}{Z_{+}}
\newcommand{\exratebuy}{Z_{-}}
\newcommand{\deltasell}{\Delta^{+}}
\newcommand{\deltabuy}{\Delta^{-}}
\newcommand{\intsell}{\lambda^{+}}
\newcommand{\intbuy}{\lambda^{-}}

\newcommand{\depth}{p^{2}}
\newcommand{\expdecay}{k}
\newcommand{\oracleprice}{s}
\newcommand{\oraclepricestochastic}{S_{t}}
\newcommand{\ppbuy}{N^{-,\feeLTbuys}}
\newcommand{\ppsell}{N^{+,\feeLTsells}}
\newcommand{\pencons}{\phi}

\newcommand{\Cash}{\mathfrak{C}}
\newcommand{\cash}{\mathfrak{c}}